\documentstyle[preprint,eqsecnum,aps]{revtex}
\begin{document}
\draft
\preprint{{\bf ROME1-1095/95}, hep-ph/9508237}
\title{
Neutralino Decays in the Minimal \protect\\
Supersymmetric Standard Model}
\author{Sandro Ambrosanio\thanks{
{\it e-mail:} {\tt ambrosanio@roma1.infn.it}}
and Barbara Mele\thanks{
{\it e-mail:} {\tt mele@roma1.infn.it}}}
\address{Dipartimento di Fisica, Universit\`a ``La Sapienza''
and I.N.F.N., Sezione di Roma, \protect\\
P.le Aldo Moro 2, I-00185 Rome, Italy}
\date{August 4, 1995}
\maketitle
\begin{abstract}
A complete phenomenological study of the next-to-lightest
neutralino decays is performed in the MSSM.
The widths and branching ratios for all the possible
decay channels (including the radiative decay
$\tilde{\chi}^{\scriptscriptstyle  0}_2  \rightarrow
\tilde{\chi}^{\scriptscriptstyle  0}_1  \gamma$
and the decay into a light Higgs
$\tilde{\chi}^{\scriptscriptstyle  0}_2  \rightarrow
\tilde{\chi}^{\scriptscriptstyle  0}_1  h^{\scriptscriptstyle  0} $)
are studied in detail as functions of all the {\sc SuSy} parameters
of the model.
Particular attention is paid to situations that are interesting for
LEP2 searches. Non-trivial decay patterns are found that critically
depend on the region of the parameter space considered. \protect\\
{\bf Pacs numbers:} 14.80.Ly, 12.60.Jv
\end{abstract}
\thispagestyle{empty}

\vspace*{\fill}

\begin{center}
{\large Submitted to {\it Physical Review D}}
\end{center}

\newpage
\thispagestyle{empty}
\null
\newpage

\setcounter{page}{1}

\section{Introduction}
\label{sec:intro}

The introduction of supersymmetry ({\sc SuSy}) can solve the hierarchy
problems in the Standard Model (SM) only if {\sc SuSy} is broken at
the TeV scale. This implies that the {\sc SuSy} partners of the known
particles should be produced at $e^{\scriptscriptstyle
+}e^{\scriptscriptstyle  -} $\ and $pp$ collider machines planned for
the next years. The possibility of observing the new states depends
not only on their production cross sections but also on their
particular decays and consequent signatures, that might or might not
allow their detection in real experiments. Hence, a complete knowledge
of the decay structure and relevant branching ratios (BR's) of the
lightest {\sc SuSy} states (the first that could be detected) is
crucial for discussing the discovery potential of the different
machines.

In the Minimal Supersymmetric Standard Model (MSSM) \cite{hk}, among
the lightest particles in the {\sc SuSy} spectrum, there are 4
neutralinos (the {\sc SuSy} partners of the neutral electroweak (EW)
gauge and Higgs bosons) and 2 charginos (the partners of the charged
gauge and Higgs bosons). In most scenarios, apart from the Lightest
{\sc SuSy} Particle (LSP), which is in general assumed to be the
lightest neutralino ($\tilde{\chi}^{\scriptscriptstyle  0}_1 $)
(stable and invisible), the particles that could be first observed at
future experiments are the next-to-lightest neutralino
($\tilde{\chi}^{\scriptscriptstyle  0}_2 $) and the light chargino
($\tilde{\chi}^{\scriptscriptstyle  \pm}_1 $). In particular, the
production of $\tilde{\chi}^{\scriptscriptstyle  0}_1
\tilde{\chi}^{\scriptscriptstyle  0}_2 $ pairs at
$e^{\scriptscriptstyle  +}e^{\scriptscriptstyle  -} $ colliders could
allow the study of a wide region of the {\sc SuSy} parameter space
\cite{amb-mele}. To this respect, it is crucial to know as well as
possible the decay characteristics of the
$\tilde{\chi}^{\scriptscriptstyle  0}_2 $, that determine the features
of the observed signal.

Analytical results for the neutralino decay widths have been
thoroughly studied in \hbox{Refs.~}{}\cite{bartl}-\cite{hab-wyl}.
Nevertheless, at the present time, a complete phenomenological
analysis, that investigates the different kinematical and dynamical
features of neutralino decays corresponding to different regions of
the {\sc SuSy} parameter space, is still missing to our knowledge.

In this paper, we present a comprehensive study of the partial decays
widths and BR's (including the radiative decay
$\tilde{\chi}^{\scriptscriptstyle  0}_2 \rightarrow
\tilde{\chi}^{\scriptscriptstyle  0}_1 \gamma$ and the decay into a
light Higgs $\tilde{\chi}^{\scriptscriptstyle  0}_2  \rightarrow
\tilde{\chi}^{\scriptscriptstyle  0}_1  h^{\scriptscriptstyle  0} $)
of the next-to-lightest neutralino in the MSSM. The dependence on all
the {\sc SuSy} parameters is carefully considered, and non-trivial
behaviours are found when varying the different parameters.

We assume the usual MSSM framework \cite{hk}, that is: \\ 1) Minimal
content of particles and gauge groups, \\ 2) Unification conditions
for gauge couplings, gaugino and scalar masses at the GUT
(Grand-Unification Theory) scale, \\ 3) $R$-parity conserved.

We also assume that the lightest neutralino is the LSP.

All masses and couplings are set by choosing the values of a finite
set of parameters at the GUT scale: $m_0$ (the common scalar mass),
$m_{1/2}$ (the common gaugino mass), $\mu$ (the {\sc SuSy}
Higgs-mixing mass) and $\tan \beta $ (the ratio of vacuum expectation
values for the two Higgs doublets). A further parameter,
$m_{A^{\scriptscriptstyle  0}} $, is needed to describe the Higgs
sector, in case one does not use the constraints coming from the
requirement that the radiative electroweak-symmetry breaking take
place at the correct scale.

In Appendices A and B, we describe the equations that allow to get the
complete {\sc SuSy} mass spectrum and couplings starting from the
above parameters in a standard approximation. We neglect the
possibility of mixing between left and right scalar partners of
fermions, that can be  relevant in the top-stop sector, since this has
a marginal role in our study. As for the Higgs sector (that is
composed by two minimal doublets), we include the leading logarithmic
radiative corrections to masses and couplings, as described in
Appendix B.

The present work complements \hbox{Ref.~}{}\cite{amb-mele}, where
$\tilde{\chi}^{\scriptscriptstyle  0}_1
\tilde{\chi}^{\scriptscriptstyle  0}_2 $ production rates and
signatures have been studied at LEP2, by studying extensively the
decay features of the $\tilde{\chi}^{\scriptscriptstyle  0}_2 $ for a
wide choice of {\sc SuSy} parameters. Particular attention is paid to
scenarios that are typical of LEP2 physics.

The plan of the paper is the following. In section 2, all the
next-to-lightest neutralino decay channels in the MSSM are reviewed.
Also, we study contour plots of the neutralino-neutralino and
neutralino-chargino mass differences, that are crucial for the
analysis of the kinematical features of the decays. In section 3,
neutralino BR's are presented in the $(\mu, M_2)$ plane. In section 4,
some specific scenarios, that are of interest for LEP2 neutralino
searches, are analysed. In section 5, the hypothesis of a light Higgs
boson is considered. Finally, in section 6, the radiative decay
$\tilde{\chi}^{\scriptscriptstyle  0}_2  \rightarrow
\tilde{\chi}^{\scriptscriptstyle  0}_1  \gamma$ is studied. In
Appendices A and B, as anticipated above, the neutralino and chargino
mass matrices and the scalar-sector mass spectrum are discussed,
respectively.

\section{Neutralino-decay classification}
\label{sec:class}

In the MSSM, four fermionic partners of the neutral components of the
SM electroweak gauge and Higgs bosons are predicted: the photino
$\tilde{\gamma} $, the Z-ino $\tilde{Z} $ (mixtures of the U(1)
$\tilde{B}$ and SU(2) $\tilde{W}_3$ gauginos), and the two higgsinos
$\tilde{H}^{\scriptscriptstyle  0}_1 $ and
$\tilde{H}^{\scriptscriptstyle  0}_2 $ (partners of the two
Higgs-doublet neutral components). In general, this interaction
eigenstates mix, their mixing being controlled by a mass matrix $Y$
(see Appendix A). By solving a 4-th degree eigenvalue equation, one
can find the expressions of $m_{\tilde{\chi}^0_i} $ ($i=1,\ldots,4$)
and of the physical composition of the corresponding eigenstates in
terms of the set of independent parameters $\mu$, $M_2$ and $\tan
\beta $. Here, we are mainly concerned with the two lightest
neutralino states ($i =1,2$). The best direct experimental limits on
the $\tilde{\chi}^{\scriptscriptstyle  0}_1 $ and
$\tilde{\chi}^{\scriptscriptstyle  0}_2 $ masses exclude the ranges
$m_{\tilde{\chi}^0_1}  < 20 {\rm\,GeV} $ and $m_{\tilde{\chi}^0_2}  <
46 {\rm\,GeV} $, under the assumption that $\tan \beta  > 2$, at LEP1.
These limits disappear if $\tan \beta  > 1.6$ \cite{L3}. At LEP2, due
to the smaller relative importance of the $Z^{\scriptscriptstyle  0}
$-exchange diagram in the $\tilde{\chi}^{\scriptscriptstyle  0}_1
\tilde{\chi}^{\scriptscriptstyle  0}_2 $ production, different
physical components of neutralinos (and not only higgsinos) come into
play and the common scalar mass $m_0$ becomes a relevant parameter
too. In this framework, in order to put new direct limits on the
neutralino masses, one must have a complete knowledge also of the
$\tilde{\chi}^{\scriptscriptstyle  0}_2 $ decay pattern.

In \hbox{Ref.~}{}\cite{amb-mele}, the behaviour of the
$\tilde{\chi}^{\scriptscriptstyle  0}_{1,2} $ gaugino and higgsino
components is studied in detail in the {\sc SuSy} parameter space.
This is crucial also to understand the dynamics of the neutralino
decays, since different components are coupled to different particles.
For instance, in the tree level decays of neutralinos
$\tilde{\chi}^{\scriptscriptstyle  0}_i  \rightarrow
\tilde{\chi}^{\scriptscriptstyle  0}_j  f\bar{f} $, there are two main
contributions coming from the $Z^{\scriptscriptstyle  0} $ and
sfermion exchanges. While the gaugino components couple to the
scalars, the higgsino components couple only to the
$Z^{\scriptscriptstyle  0} $ boson, with different strength (in the
$m_f = 0$ limit). Also the neutralino mass spectrum depends on the
same three parameters $\mu$, $M_2$ and $\tan \beta $. A detailed
discussion on the $\tilde{\chi}^{\scriptscriptstyle  0}_i $ mass
spectrum can be found in \hbox{Refs.~}{}\cite{amb-mele} and
\cite{bartl89}.

In what follows, we list all the possible next-to-lightest neutralino
decays in the MSSM. In general, these channels are valid also for
heavier neutralinos, although the possibility of cascade decays can
make the decay structure of the heavier neutralinos more complicated.
\begin{description}
\item[a)] Decay into charged leptons:
\begin{equation}
\tilde{\chi}^{\scriptscriptstyle  0}_2 \rightarrow
\tilde{\chi}^{\scriptscriptstyle  0}_1
e^{\scriptscriptstyle  +}e^{\scriptscriptstyle  -}  \; ,
\label{n2ton1ee}
\end{equation}
or $e^{\scriptscriptstyle  \pm}  \rightarrow
\mu^{\scriptscriptstyle  \pm} , \; \tau^{\scriptscriptstyle  \pm} $;
\item[b)] decay into a neutrino pair:
\begin{equation}
\tilde{\chi}^{\scriptscriptstyle  0}_2  \rightarrow
\tilde{\chi}^{\scriptscriptstyle  0}_1
\nu_{\scriptscriptstyle  \ell} \bar{\nu}_{\scriptscriptstyle  \ell}   \; ,
\label{n2ton1vv}
\end{equation}
where $\ell = e, \mu, \tau$;
\item[c)] decay into a light-quark pair:
\begin{equation}
\tilde{\chi}^{\scriptscriptstyle  0}_2  \rightarrow
\tilde{\chi}^{\scriptscriptstyle  0}_1  q\bar{q}  \; ,
\label{n2ton1qq}
\end{equation}
where $q=u,d,s,c,b$;
\item[d)] cascade decay through a real chargino:
\begin{equation}
\begin{array}{r c l}
\tilde{\chi}^{\scriptscriptstyle  0}_2
& \rightarrow &
f_1 \bar{f}_1^{\prime} \;
\tilde{\chi}^{\scriptscriptstyle \pm}_1 \\
& & \phantom{f_1 \bar{f}_1^{\prime} \;}
\:\raisebox{1.3ex}{\rlap{$\vert$}}\!\rightarrow
f_2 \bar{f}_2^{\prime}
\tilde{\chi}^{\scriptscriptstyle  0}_1  \; ,
\end{array}
\label{eq:cascade}
\end{equation}
where each pair of fermions $f_i f_i^{\prime}$ in the final state
is an isospin doublet of either leptons or light quarks;
\item[e)] decay into a light scalar ($h^{\scriptscriptstyle  0} $)
or pseudoscalar ($A^{\scriptscriptstyle  0} $) Higgs:
\begin{mathletters}
\label{ntoh}
\begin{eqnarray}
\tilde{\chi}^{\scriptscriptstyle  0}_2
& \rightarrow  & \tilde{\chi}^{\scriptscriptstyle  0}_1
h^{\scriptscriptstyle  0}  \label{n2ton1h} \; , \\
\tilde{\chi}^{\scriptscriptstyle  0}_2
& \rightarrow   & \tilde{\chi}^{\scriptscriptstyle  0}_1
A^{\scriptscriptstyle  0}  \label{n2ton1A} \; ,
\end{eqnarray}
\end{mathletters}
where $h^{\scriptscriptstyle  0} $ and $A^{\scriptscriptstyle  0} $
are part of the MSSM Higgs doublet;
\item[f)] radiative decay into a photon:
\begin{equation}
\tilde{\chi}^{\scriptscriptstyle  0}_2  \rightarrow
\tilde{\chi}^{\scriptscriptstyle  0}_1  \gamma \; .
\label{n2ton1ph}
\end{equation}
\end{description}

The first three channels occur through either a $Z^{\scriptscriptstyle
 0} $ or a scalar-particle exchange. Different scalar partners come
into play: (left or right) selectron (in channel {\bf a}), (left)
sneutrino (in channel {\bf b}) and (left or right) squark (in channel
{\bf c}) (see \hbox{Fig.~}{}\ref{fey3bneut}). Assuming massless
fermions makes the channels proceeding through neutral Higgs bosons
vanish in {\bf a}, {\bf b} and {\bf c}. We name $s$-channels the
contributions from diagrams with the two neutralinos entering the same
vertex ($Z^{\scriptscriptstyle  0} $ exchange), and ($t,u$)-channels
the ones where the two neutralinos enter different vertices (sfermion
exchange).

Whenever the $\tilde{\chi}^{\scriptscriptstyle  0}_2 $ is heavier than
some scalar fermions, the corresponding channels will proceed through
two steps via real sparticles.

A possible gluino in the final state
($\tilde{\chi}^{\scriptscriptstyle  0}_2  \rightarrow   \tilde{g}
q\bar{q} $) is excluded by the gaugino mass unification hypothesis,
that makes gluinos considerably heavier than light neutralinos.

Cascade decays through a real chargino ({\bf d}) occur via similar
graphs (\hbox{Fig.~}{}\ref{fey3bcasc}). The diagrams for the
second-step decay (\hbox{cfr.}{}\ \hbox{Eq.~}{}(\ref{eq:cascade})) can
be obtained by the same graphs by exchanging the neutralino and the
chargino.

As for the channel {\bf e}, there are five possible Higgses (either
neutral or charged) that could contribute to the tree-level
$\tilde{\chi}^{\scriptscriptstyle  0}_2 $ decays into a scalar Higgs
boson plus a light neutralino or chargino. For the next-to-lightest
neutralino, only decays into the two lightest bosons (\hbox{i.e.}{},
the lightest neutral scalar and the pseudoscalar Higgses) can be
present, for the moderate $\tilde{\chi}^{\scriptscriptstyle  0}_2 $
masses we are considering here (\hbox{Fig.~}{}\ref{fey2bhiggs}).

One important point to keep in mind in the decay study is that,
whenever the $\tilde{\chi}^{\scriptscriptstyle  0}_2 $ can decay into
a real scalar plus a fermion (\hbox{e.g.}{}, a selectron plus an
electron or a Higgs plus a lightest neutralino), this channel tends to
saturate the corresponding width and BR. The same occurs when the mass
difference between the two lightest neutralino is sufficient to allow
the decay into a real $Z^{\scriptscriptstyle  0} $. In the latter
case, the relevant BR's for different signatures recover the
$Z^{\scriptscriptstyle  0} $ ones. However, the last possibility never
occurs in the LEP2 parameter regions.

We point out that, apart from the decays into Higgses, that are
considered only if the two-body on-shell decay is allowed by the phase
space, our treatment of the three-body decays always takes into
account properly the possibility of decays into two real particles,
whenever this is permitted.

In \hbox{Fig.~}{}\ref{n2-n1mass}, we show the contour plot for the
mass difference between $\tilde{\chi}^{\scriptscriptstyle  0}_2 $ and
$\tilde{\chi}^{\scriptscriptstyle  0}_1 $, for $\tan \beta  = 1.5$ and
30, in the $(\mu, M_2)$ plane. From these plots, one can immediately
infer, for a given scalar mass, which are the parameter regions where
the decays into real scalars are kinematically allowed, and
consequently can dominate the $\tilde{\chi}^{\scriptscriptstyle  0}_2
$ decay.

Furthermore, in \hbox{Fig.~}{}\ref{n2-c1mass}, the difference between
$m_{\tilde{\chi}^0_2} $ and $m_{\tilde{\chi}^{\pm}_1} $ is plotted.
Shaded area represent situations where this difference is negative and
the neutralino cascade decays through a chargino are not allowed. For
small $\tan \beta $, one can anticipate a sizeable BR for cascade
decays in the positive $\mu$ half-plane (\hbox{cfr.}{}\ section 3).

Diagrams contributing to the radiative decay
$\tilde{\chi}^{\scriptscriptstyle  0}_2  \rightarrow
\tilde{\chi}^{\scriptscriptstyle  0}_1 \gamma$ are shown in
\hbox{Fig.~}{}\ref{feyrad}, where the corresponding graphs with
clockwise circulating particles in the loops must be added. The fields
$G^{\scriptscriptstyle  \pm} $ are the Goldstone quanta giving masses
to charged vector bosons. We assume the nonlinear R-gauge, that is
described in \hbox{Ref.~}{}\cite{hab-wyl}. One can see that there are
many physical charged particles flowing in the loops: all charged
standard fermions and their corresponding scalar partners, the charged
vector and Higgs bosons and their fermionic partners, the charginos.
In the {\sc SuSy} parameter scheme we adopt, relevant contributions
come mostly from  the $W^{\scriptscriptstyle  \pm} $/chargino and the
top/stop loops, with non-negligible interferences. In this decay, the
visible part of the final state is given by a monochromatic photon.
{}From \hbox{Fig.~}{}\ref{n2-n1mass}, one can get information on the
final photon energy.

In the following analysis, we mainly concentrate on the MSSM parameter
regions not excluded at LEP1. Particular attention is given to regions
explorable at the forthcoming experiments, especially at LEP2 (that
are shown in \hbox{Fig.~}{}\ref{lep2scen}).

For definiteness, we restrict to the following ranges of {\sc SuSy}
parameters:
\begin{mathletters}
\label{eq:parng}
\begin{eqnarray}
0       \le
& M_2  & \le 4 M_{\scriptscriptstyle  Z}           \\
-4M_{\scriptscriptstyle  Z}    \le
& \mu  & \le 4 M_{\scriptscriptstyle  Z}           \\
40 {\rm\,GeV}  \le
& m_0  & \le 500 {\rm\,GeV}        \\
1        <  & \tan \beta
& \le 60   		\\
M_{\scriptscriptstyle  Z}      \le
& m_{A^{\scriptscriptstyle  0}}   & \le 3 M_{\scriptscriptstyle  Z}
\end{eqnarray}
\end{mathletters}
The lower limit on $m_0$ is connected to present experimental limits
on the masses of the {\sc SuSy} partners of leptons and quarks.
It generally  excludes scenarios where the LSP is a scalar.

\section{Study of the $\protect\tilde{\chi}^{\scriptscriptstyle  0}_2$
BR's in the $(\mu, M_2)$ plane}
\label{sec:BRvsmum2}

In this section, we make a detailed study of the BR's for the decay
channels {\bf a-e} (defined in the previous section) in the $(\mu,
M_2)$ plane, at fixed values of $m_0$ and $\tan \beta $. We assume
$m_{A^{\scriptscriptstyle  0}}  = 3 M_{\scriptscriptstyle  Z} $, that,
unless $\tan \beta $ is near 1, generally implies a
$h^{\scriptscriptstyle  0} $ mass above the threshold for the channel
$\tilde{\chi}^{\scriptscriptstyle  0}_2  \rightarrow
\tilde{\chi}^{\scriptscriptstyle  0}_1  h^{\scriptscriptstyle  0} $,
in the $(\mu, M_2)$ region covered by LEP2 searches. The light-Higgs
case will be considered in section 5, while in section 6 we will
concentrate on the radiative decay $\tilde{\chi}^{\scriptscriptstyle
0}_2  \rightarrow   \tilde{\chi}^{\scriptscriptstyle  0}_1  \gamma$.
Everywhere, the widths and BR's connected to the decays into charged
leptons are relative to a single species, while the decays into
neutrinos are summed over three families and the decays into quarks
over five light flavours. Analogously, for cascade decays, the BR for
the leptonic channel is for one single species, while the hadronic
channels are summed over two light quark doublets. At this stage, the
second-step decay of the $\tilde{\chi}^{\scriptscriptstyle  \pm}_1 $
is not considered.

The BR's for each channel are studied in the $(\mu, M_2)$ plane for
four different choices of the $m_0$ and $\tan \beta $ parameter,
namely:
\begin{mathletters}
\label{br_cases}
\begin{eqnarray}
{(\rm a)}: \:\:  (m_0, \tan \beta )
& = & (M_{\scriptscriptstyle  Z} , 1.5),  \\
{(\rm b)}: \:\:  \phantom{(m_0, \tan \beta )}
& = & (M_{\scriptscriptstyle  Z} , 30),   \\
{(\rm c)}: \:\:  \phantom{(m_0, \tan \beta )}
& = & (3M_{\scriptscriptstyle  Z} , 1.5), \\
{(\rm d)}: \:\:  \phantom{(m_0, \tan \beta )}
& = & (3M_{\scriptscriptstyle  Z} , 30),
\end{eqnarray}
\end{mathletters}
for $|\mu| \le 4 M_{\scriptscriptstyle  Z} $ and $0 \le M_2 \le 4
M_{\scriptscriptstyle  Z} $. The observed behaviour is in general
highly non-trivial, due to both the sharp dependence of the neutralino
physical composition on $\mu$ and $M_2$ \cite{amb-mele} and the
corresponding variation in the neutralino mass spectrum
(\hbox{cfr.}{}\ \hbox{Fig.~}{}\ref{n2-n1mass}). Also a weak dependence
on $M_2$ comes from the spectrum of the scalar masses that enter the
$t$-channel contributions to the $\tilde{\chi}^{\scriptscriptstyle
0}_2 $ decay (\hbox{Fig.~}{}\ref{fey3bneut}).

In \hbox{Fig.~}{}\ref{brnee}, we present the BR for the decay
$\tilde{\chi}^{\scriptscriptstyle  0}_2 \rightarrow
\tilde{\chi}^{\scriptscriptstyle  0}_1  e^{\scriptscriptstyle
+}e^{\scriptscriptstyle  -} $. In order to get large BR's, in this
case, one needs relatively small $m_0$ values, so that the $t$-channel
contributions get substantial with respect to the
$Z^{\scriptscriptstyle  0} $-exchange channel. In this way, one
obtains leptonic BR's much larger than the corresponding
BR($Z^{\scriptscriptstyle  0} \rightarrow  \ell^{\scriptscriptstyle
+}\ell^{\scriptscriptstyle  -} $). For instance, when $m_0 =
M_{\scriptscriptstyle  Z} $ (\hbox{Fig.~}{}\ref{brnee}a,b), one always
gets wide regions of the plane where
BR($\tilde{\chi}^{\scriptscriptstyle  0}_1  e^{\scriptscriptstyle
+}e^{\scriptscriptstyle  -} $) is larger than 25\% (that corresponds
to a BR $\ge 75\%$ when summed over three lepton species). However,
when considering the LEP2 realm (\hbox{Fig.~}{}\ref{lep2scen}), the
relative importance of these regions is reduced, especially for large
$\tan \beta $ values. Note that at large $\tan \beta $, the behaviour
tends to be more symmetric with respect to the line $\mu = 0$
(\hbox{Fig.~}{}\ref{brnee}b,d).

In order to guarantee the detectability of the
$\tilde{\chi}^{\scriptscriptstyle  0}_1 e^{\scriptscriptstyle
+}e^{\scriptscriptstyle  -} $ final state in ordinary collider
experiments, it is also useful to consider a threshold on the minimum
energy for an observable $e^{\scriptscriptstyle
+}e^{\scriptscriptstyle  -} $ state. The effect of this condition can
be guessed through \hbox{Fig.~}{}\ref{n2-n1mass}, since
$(m_{\tilde{\chi}^0_2}  - m_{\tilde{\chi}^0_1} )$ is directly
connected to the final $e^{\scriptscriptstyle  +}e^{\scriptscriptstyle
 -} $ energy. For instance, one can see that the rejection of the
areas where $(m_{\tilde{\chi}^0_2}  - m_{\tilde{\chi}^0_1} ) < 20
{\rm\,GeV}  $ has a moderate influence on LEP2 physics (\hbox{cfr.}{}\
\hbox{Fig.~}{}\ref{lep2scen}).

In \hbox{Fig.~}{}\ref{brnee}a, the large BR at moderate values of
$M_2$ and $\mu < 0$ is due to the opening of the tree-level channel
$\tilde{\chi}^{\scriptscriptstyle  0}_2 \rightarrow
e^{\scriptscriptstyle  \mp} \tilde{e}^{\scriptscriptstyle
\pm}_{\scriptscriptstyle  R} $ at $m_0 \approx M_{\scriptscriptstyle
Z} $, which is not contrasted by $\tilde{\chi}^{\scriptscriptstyle
0}_2  \rightarrow   \nu_{\scriptscriptstyle  e}
\tilde{\nu}_{\scriptscriptstyle  e,L} $ (in general $m_{\tilde{e}_R} <
m_{\tilde{\nu}_{e,L}} $, \hbox{cfr.}{}\ Appendix B). See
\hbox{Fig.~}{}\ref{wbrvsm0a} for further details.

In \hbox{Fig.~}{}\ref{brnvv}, the channel
$\tilde{\chi}^{\scriptscriptstyle  0}_2  \rightarrow
\tilde{\chi}^{\scriptscriptstyle  0}_1  \nu_{\scriptscriptstyle  \ell}
\bar{\nu}_{\scriptscriptstyle  \ell}  $ is studied. As in the previous
case, low $m_0$ values tend to enhance the BR. Note that, in both the
charged lepton and the neutrino case, the corresponding
$Z^{\scriptscriptstyle  0} $ BR's are recovered in the region of small
$|\mu|$ and $M_2 \;\raisebox{-.5ex}{\rlap{$\sim$}}
\raisebox{.5ex}{$>$}\;$ (1-2) $M_{\scriptscriptstyle  Z} $. Indeed, in
this region, the higgsino components and, consequently, the
$Z^{\scriptscriptstyle  0} $-exchange channel are dominant,
independently of $m_0$ and $\tan \beta $.

This is also true for the hadronic channel that is considered in
\hbox{Fig.~}{}\ref{brnqq}. On the other hand, in the hadronic decay, a
low $m_0$ value can decrease the BR with respect to the
$Z^{\scriptscriptstyle  0} $-channel expectation. This is due, for a
given $m_0$, to the larger value of the squark masses (entering the
$t$-channel contribution), compared to the slepton masses
(\hbox{cfr.}{}\ Appendix B). For large $m_0$, $t$-channels tend to
vanish, and the BR for $\tilde{\chi}^{\scriptscriptstyle  0}_2
\rightarrow \tilde{\chi}^{\scriptscriptstyle  0}_1  f\bar{f} $
recovers the $Z^{\scriptscriptstyle  0}  \rightarrow   f\bar{f} $ one.

Cascade-decay BR's are shown in \hbox{Figs.~}{}\ref{brclv} and
\ref{brcqq} for the channels $\tilde{\chi}^{\scriptscriptstyle  0}_2
\rightarrow   \tilde{\chi}^{\scriptscriptstyle  \pm}_1
e^{\scriptscriptstyle  \mp}  \nu_{\scriptscriptstyle  e} $ and
$\tilde{\chi}^{\scriptscriptstyle  0}_2  \rightarrow   \sum_q
\tilde{\chi}^{\scriptscriptstyle  \pm}_1  q \bar{q}^{\prime}$,
respectively. At small values of $\tan \beta $, the importance of this
channel is restricted to the positive-$\mu$ half-plane in connection
to the regions where $\tilde{\chi}^{\scriptscriptstyle  \pm}_1 $ is
sufficiently lighter than $\tilde{\chi}^{\scriptscriptstyle  0}_2 $
(\hbox{cfr.}{}\ \hbox{Figs.~}{}\ref{brclv}a,c, \ref{brcqq}a,c and
\ref{n2-c1mass}). The leptonic channel can reach a BR of 10\% for each
leptonic species at low $\tan \beta $. The relevance of this decay is
further increased at larger $m_0$ (\hbox{cfr.}{}\
\hbox{Fig.~}{}\ref{brclv}c). Raising $\tan \beta $ makes the pattern
symmetrical with respect to the $\mu = 0$ axis, although the BR never
reaches a sizeable level in the interesting regions (\hbox{cfr.}{}\
\hbox{Figs.~}{}\ref{brclv}b,d). An analogous situation is observed for
the decay into $\tilde{\chi}^{\scriptscriptstyle  \pm}_1 q
\bar{q}^{\prime}$ in \hbox{Fig.~}{}\ref{brcqq}.

In \hbox{Fig.~}{}\ref{brnlh} we study the channel
$\tilde{\chi}^{\scriptscriptstyle  0}_2  \rightarrow
\tilde{\chi}^{\scriptscriptstyle  0}_1  h^{\scriptscriptstyle  0} $.
Whenever the light-Higgs mass is lighter than the difference
$(m_{\tilde{\chi}^0_2}  - m_{\tilde{\chi}^0_1} )$, this process has a
large BR due to the two-body nature of the decay. The Higgs mass
spectrum is fixed here by $m_{A^{\scriptscriptstyle  0}}  = 3
M_{\scriptscriptstyle  Z} $, that corresponds to
$m_{h^{\scriptscriptstyle  0}} $ in the range $ 50 \div  90 {\rm\,GeV}
$, at $\tan \beta  = 1.5$ and $100 \div 130 {\rm\,GeV} $, at $\tan
\beta  = 30$, for the assumed range of $m_0$ and $M_2$ (\hbox{cfr.}{}\
Appendix B).

The most favourable case for the process
$\tilde{\chi}^{\scriptscriptstyle  0}_2  \rightarrow
\tilde{\chi}^{\scriptscriptstyle  0}_1  h^{\scriptscriptstyle  0} $ is
the one with small $\tan \beta $ and $m_0$ (\hbox{cfr.}{}\
\hbox{Fig.~}{}\ref{brnlh}a). Even in this case, the decay threshold
opens mostly at relatively large values of $|\mu|$, that correspond to
heavier neutralinos. Increasing $m_0$ (\hbox{cfr.}{}\
\hbox{Fig.~}{}\ref{brnlh}c), slightly restricts the allowed regions,
due to the $m_{\tilde{t}} $ dependence of the
$m_{h^{\scriptscriptstyle  0}} $ radiative corrections (\hbox{cfr.}{}\
Appendix B), but, at the same time, increases the branching fraction,
due to the depletion of all the other
$\tilde{\chi}^{\scriptscriptstyle  0}_2  \rightarrow
\tilde{\chi}^{\scriptscriptstyle  0}_1  f\bar{f} $ channels. At $\tan
\beta  = 30$, the decay $\tilde{\chi}^{\scriptscriptstyle  0}_2
\rightarrow   \tilde{\chi}^{\scriptscriptstyle  0}_1
h^{\scriptscriptstyle  0} $ is not allowed in almost the whole $(\mu,
M_2)$ plane considered, due to the increase of
$m_{h^{\scriptscriptstyle  0}} $ with $\tan \beta $ (\hbox{cfr.}{}\
Appendix B) and also to the decrease of $(m_{\tilde{\chi}^0_2}  -
m_{\tilde{\chi}^0_1} )$ (\hbox{cfr.}{}\
\hbox{Fig.~}{}\ref{n2-n1mass}). In \hbox{Figs.~}{}\ref{brnlh}b,d an
intermediate-$\tan \beta $ situation is shown.

Note that the case of a lighter $h^{\scriptscriptstyle  0} $ (or
$m_{A^{\scriptscriptstyle  0}} $) can considerably alter the pattern
of the $\tilde{\chi}^{\scriptscriptstyle  0}_2 $ BR's. The case of a
lighter Higgs will be considered in section 5.

\section{Neutralino decays at LEP2}
\label{sec:LEP2dec}

In our analysis of the production of $\tilde{\chi}^{\scriptscriptstyle
 0}_1 \tilde{\chi}^{\scriptscriptstyle  0}_2 $ pairs at LEP2 in
\hbox{Ref.~}{}\cite{amb-mele}, we have identified particular regions
and scenarios in the parameter space. These are characterized by
specific dynamical and kinematical properties of the light
neutralinos. We have defined as NR$^{\rm \pm}$ ({\it Neutralino
Regions}) the areas of the $(\mu, M_2)$ plane where:
\begin{equation}
m_{\tilde{\chi}^0_1}  + m_{\tilde{\chi}^0_2}
< \sqrt{s}  < m_{\tilde{\chi}^{\pm}_1}  \; ,
\label{NR}
\end{equation}
at fixed $\tan \beta $, where $\sqrt{s} $ is the \hbox{c.m.}{}\
collision energy at LEP2 ($\sqrt{s}  \simeq 190 {\rm\,GeV} $)
(\hbox{Fig.~}{}\ref{lep2scen}). In this regions, chargino pair
production is kinematically forbidden, while, for moderate values of
scalar masses, the $\tilde{\chi}^{\scriptscriptstyle  0}_1
\tilde{\chi}^{\scriptscriptstyle  0}_2 $ production can have sizeable
cross sections. Of course, neutralino-pair production can be of help
also below the Neutralino Regions, where it complements chargino
production. On the other hand, the largest rates for
$e^{\scriptscriptstyle  +}e^{\scriptscriptstyle  -}  \rightarrow
\tilde{\chi}^{\scriptscriptstyle  0}_1
\tilde{\chi}^{\scriptscriptstyle  0}_2 $ arise for $|\mu|
\;\raisebox{-.5ex}{\rlap{$\sim$}} \raisebox{.5ex}{$<$}\;
M_{\scriptscriptstyle  Z} $ and $ M_2
\;\raisebox{-.5ex}{\rlap{$\sim$}} \raisebox{.5ex}{$>$}\;
M_{\scriptscriptstyle  Z} $ (what we call HCS$^{\rm \pm}$, that stands
for {\it High Cross Section} regions), although, in these zones,
chargino production is also allowed. Here, the higgsino components of
$\tilde{\chi}^{\scriptscriptstyle  0}_1 $ and
$\tilde{\chi}^{\scriptscriptstyle  0}_2 $ are dominant and the main
production mechanism is through $Z^{\scriptscriptstyle  0} $ exchange.

In these regions, for $\tan \beta  = 1.5$, we have chosen a set of six
specific points (shown in \hbox{Fig.~}{}\ref{lep2scen}a), that can
schematize the spectrum of possibilities for the neutralino couplings
and masses: scenarios A, B, C and D in the Neutralino Regions and
H$^{\rm \pm}$ in the High Cross Section regions. In
\hbox{Tab.~}{}\ref{tab:scentgb15}, we show the masses and physical
components of the two lightest neutralinos and the mass spectrum of
charginos and heavier neutralinos, corresponding to these points in
the $(\mu, M_2)$ plane. Moreover, we show the sfermion spectrum (that
also influences the decay properties of neutralinos) corresponding to
these cases for $m_0 = M_{\scriptscriptstyle  Z} $. A more detailed
analysis of the dynamical characteristics of these scenarios can be
found in \hbox{Ref.~}{}\cite{amb-mele}.

In this section, we present a study of the
$\tilde{\chi}^{\scriptscriptstyle  0}_2 $ widths and BR's in these
specific cases. Although the values $\tan \beta  = 1.5$ (associated to
the definition of such scenarios) and $m_0 = M_{\scriptscriptstyle  Z}
$ correspond to particularly favourable cases for neutralino
production rates, we also study the behaviour of decay widths and BR's
in a large range of $\tan \beta $ and $m_0$ values. We will call
$\tilde{A}$-$\tilde{H}^{\pm}$ the scenarios with the same values of
$\mu$ and $M_2$ as A-H$^{\rm \pm}$, but $\tan \beta  \ne 1.5$ (see,
\hbox{e.g.}{}, \hbox{Fig.~}{}\ref{lep2scen}b where $\tan \beta  =
30$).

If not otherwise specified, we will assume that decays into real Higgs
bosons are not kinematically allowed. In this case, there is some
influence of the Higgs sector only in the radiative channel.
Accordingly, the results presented in this section are obtained for
$m_{A^{\scriptscriptstyle  0}}  = 3 M_{\scriptscriptstyle  Z} $. The
effect of varying $m_{A^{\scriptscriptstyle  0}} $ will be discussed
in section 5.

In \hbox{Figs.~}{}\ref{wbrvsm0a}--\ref{wbrvsm0hp}, we present all
partial widths and BR's versus $m_0$ in the scenarios A, B, C, D,
H$^{\rm -}$ and H$^{\rm +}$. In order to understand the general
pattern of the decay widths, one must recall the specific ordering of
the scalar masses at fixed $m_0$, that is predicted by the
renormalization group equations and unification assumptions. For
instance, one always gets: $m_{\tilde{q}_{L,R}}  >
m_{\tilde{\ell}_{L,R}} $ and $m_{\tilde{f}_L}   > m_{\tilde{f}_R} $.
When $m_0$ is sufficiently small, as to allow
$\tilde{\chi}^{\scriptscriptstyle  0}_2 $ decays into one or more real
scalars, the largest decay widths are associated to the corresponding
channels, that in general are the leptonic ones. For instance, this
occurs for $m_0 \;\raisebox{-.5ex}{\rlap{$\sim$}}
\raisebox{.5ex}{$<$}\; 100 {\rm\,GeV} $, in
\hbox{Fig.~}{}\ref{wbrvsm0a}, where we study the scenario A
($\tilde{\chi}^{\scriptscriptstyle  0}_{1,2} $ dominantly gauginos,
\hbox{cfr.}{}\ \hbox{Tab.~}{}\ref{tab:scentgb15}). One can see that
the largest rates (up to $1 \div 100 {\rm\,MeV} $) are by far those
corresponding to decays into real sleptons. The width for
$\tilde{\chi}^{\scriptscriptstyle  0}_2  \rightarrow
\nu_{\scriptscriptstyle  \ell} \bar{\tilde{\nu}}_{\scriptscriptstyle
\ell,L} \rightarrow   \nu_{\scriptscriptstyle  \ell}
\bar{\nu}_{\scriptscriptstyle \ell} \tilde{\chi}^{\scriptscriptstyle
0}_1$ is summed over all neutrino flavours, contrary to the
charged-lepton channel. For $m_0 \;\raisebox{-.5ex}{\rlap{$\sim$}}
\raisebox{.5ex}{$<$}\; 60 {\rm\,GeV} $, all sleptons are produced on
the mass-shell. By increasing $m_0$, one meets, in order, the
$m_{\tilde{\ell}_L} $, $m_{\tilde{\nu}_{\ell,L}} $ and
$m_{\tilde{\ell}_R} $ thresholds. At large $m_0$, only the
$Z^{\scriptscriptstyle  0} $-exchange contributions survive. Note also
the peaks of the leptonic cascade-decay width, that correspond to the
chain of on-shell decays $\tilde{\chi}^{\scriptscriptstyle  0}_2
\rightarrow \tilde{e}^{\scriptscriptstyle  \pm}_{\scriptscriptstyle
L} e^{\scriptscriptstyle  \mp}  \rightarrow
\tilde{\chi}^{\scriptscriptstyle  \pm}_1 \nu_{\scriptscriptstyle  e}
e^{\scriptscriptstyle  \mp}$ and $\tilde{\chi}^{\scriptscriptstyle
0}_2  \rightarrow \tilde{\nu}_{\scriptscriptstyle  e,L}
\nu_{\scriptscriptstyle  e} \rightarrow
\tilde{\chi}^{\scriptscriptstyle  \pm}_1 e^{\scriptscriptstyle  \mp}
\nu_{\scriptscriptstyle  e}$. When $m_0
\;\raisebox{-.5ex}{\rlap{$\sim$}} \raisebox{.5ex}{$>$}\; 100
{\rm\,GeV} $, no two-body decay is allowed. On the other hand, in the
parameter range considered, one has in general $m_{\tilde{q}_{L,R}}  >
m_{\tilde{\chi}^0_2} $ and the width for the
$\tilde{\chi}^{\scriptscriptstyle  0}_2  \rightarrow   q\bar{q}
\tilde{\chi}^{\scriptscriptstyle  0}_1 $ is relatively small and
almost constant (between 0.1 and 1 KeV) when varying $m_0$. Concerning
the radiative decay, the curves are obtained through the complete
results of \hbox{Ref.~}{}\cite{hab-wyl}. The width for
$\tilde{\chi}^{\scriptscriptstyle  0}_2  \rightarrow
\tilde{\chi}^{\scriptscriptstyle  0}_1 \gamma$ never exceeds 1 KeV for
$m_0 \;\raisebox{-.5ex}{\rlap{$\sim$}} \raisebox{.5ex}{$>$}\;
M_{\scriptscriptstyle  Z} $.

The corresponding BR pattern closely reflects the scalar-mass
threshold structure (\hbox{Fig.~}{}\ref{wbrvsm0a}). Indeed, for $m_0
\;\raisebox{-.5ex}{\rlap{$\sim$}} \raisebox{.5ex}{$<$}\; 75 {\rm\,GeV}
$, the BR for the invisible channel $\tilde{\chi}^{\scriptscriptstyle
0}_2 \rightarrow  \sum_{\ell}\nu_{\scriptscriptstyle  \ell}
\bar{\nu}_{\scriptscriptstyle  \ell}
\tilde{\chi}^{\scriptscriptstyle  0}_1 $ is more than 80\%. Hence, in
this regime, the next-to-lightest neutralino is phenomenologically
almost equivalent to the LSP, which means that most of the times it
just produces missing energy and momentum in the final states. This
effect tends to concern even larger ranges of $m_0$ when $\tan \beta $
increases, due to the relative lowering of the sneutrino mass with
respect to the charged-slepton masses. Also, notice that the presence
of two close thresholds for the decay into a real
$\tilde{\nu}_{\scriptscriptstyle  \ell,L} $ and a real
$\tilde{\ell}_{\scriptscriptstyle  R} $ gives rise to a peak structure
in the BR($\tilde{\chi}^{\scriptscriptstyle  0}_2 \rightarrow
\ell^{\scriptscriptstyle  +}\ell^{\scriptscriptstyle  -}
\tilde{\chi}^{\scriptscriptstyle  0}_1 $) at $m_0 \simeq 83 {\rm\,GeV}
$. This corresponds to a fast deepening in the invisible BR. For $80
{\rm\,GeV}  \;\raisebox{-.5ex}{\rlap{$\sim$}} \raisebox{.5ex}{$<$}\;
m_0 \;\raisebox{-.5ex}{\rlap{$\sim$}} \raisebox{.5ex}{$<$}\; 200
{\rm\,GeV} $, the largest BR is that for charged leptons
($\;\raisebox{-.5ex}{\rlap{$\sim$}} \raisebox{.5ex}{$>$}\; 60\%$, for
all the three lepton species). For $m_0
\;\raisebox{-.5ex}{\rlap{$\sim$}} \raisebox{.5ex}{$>$}\; 200
{\rm\,GeV} $, the hadronic channel gets more and more important. It
reaches 80\% for $m_0 \simeq \frac{1}{2} {\rm\,TeV} $. The BR for the
radiative decay grows with $m_0$, although at $m_0 \simeq 500
{\rm\,GeV} $ one still has only BR$\simeq 4\%$. Anyhow, concerning
searches at LEP2, one has to keep in mind that, for $m_0
\;\raisebox{-.5ex}{\rlap{$\sim$}} \raisebox{.5ex}{$>$}\; 300
{\rm\,GeV} $, the production rate for
$\tilde{\chi}^{\scriptscriptstyle  0}_1
\tilde{\chi}^{\scriptscriptstyle  0}_2 $ pairs is below the
detectability threshold for a realistic machine luminosity (see
\hbox{Ref.~}{}\cite{amb-mele}).

In \hbox{Fig.~}{}\ref{wbrvsm0b}, we deal with the scenario B, where
the $\tilde{\chi}^{\scriptscriptstyle  0}_1 $ is mainly a photino and
the $\tilde{\chi}^{\scriptscriptstyle  0}_2 $ is mainly a
$\tilde{H}^{\scriptscriptstyle  0}_b $ (\hbox{cfr.}{}\
\hbox{Tab.~}{}\ref{tab:scentgb15}). In general, the behaviour of
widths and BR's is qualitatively similar to the ones in scenario A,
apart from the absence of the decay into
$\tilde{\ell}_{\scriptscriptstyle  L} $ (whose mass is above
threshold) and the presence of rather strong destructive interference
effects in the leptonic channels. The latter are clearly visible for
the process $\tilde{\chi}^{\scriptscriptstyle  0}_2  \rightarrow
\tilde{\chi}^{\scriptscriptstyle  0}_1  \nu_{\scriptscriptstyle  \ell}
\bar{\nu}_{\scriptscriptstyle  \ell}  $ around $m_0 \simeq 64
{\rm\,GeV} $ and in the case $\tilde{\chi}^{\scriptscriptstyle  0}_2
\rightarrow   \tilde{\chi}^{\scriptscriptstyle  0}_1
e^{\scriptscriptstyle  +}e^{\scriptscriptstyle  -} $ for $m_0 \simeq
130 {\rm\,GeV} $ (\hbox{Fig.~}{}\ref{wbrvsm0b}). Indeed, the different
physical nature of $\tilde{\chi}^{\scriptscriptstyle  0}_1 $ and
$\tilde{\chi}^{\scriptscriptstyle  0}_2 $ gives rise, in particular
$m_0$ ranges, to a comparable size for the $s$- and $t$-channel decay
contributions with negative interference of the same order of
magnitude. For instance, in the minimum of the width for the decay
into electrons, the $s$-channel and $t$-channel contribute 65 and 50
MeV, respectively, while their interference gives -95 MeV.

In the scenario C (\hbox{cfr.}{}\ \hbox{Tab.~}{}\ref{tab:scentgb15})
the larger value of $M_2$ with respect to the previous scenarios
generates larger masses in the scalar sector, particularly in the
left-handed sector. As a consequence, only the
$\tilde{\ell}_{\scriptscriptstyle  R} $ can go on mass-shell, while
both the invisible and the hadronic widths, that are dominated by the
$Z^{\scriptscriptstyle  0} $-exchange, keep constant
(\hbox{Fig.~}{}\ref{wbrvsm0c}). As for BR's, the charged leptonic
channels saturate the width up to $m_0 \simeq 64 {\rm\,GeV} $
(BR$\simeq 33\%$ for each lepton species)
(\hbox{Fig.~}{}\ref{wbrvsm0c}). Note that the relatively small widths
for the tree level channels enhance the BR for the radiative decay,
for $m_0 \;\raisebox{-.5ex}{\rlap{$\sim$}} \raisebox{.5ex}{$>$}\; 70
{\rm\,GeV} $, up to about 15\%.

The scenario D is chosen in the positive-$\mu$ range (contrary to the
previous ones) and, in particular, in the area of the Neutralino
Regions where $\tilde{\chi}^{\scriptscriptstyle  0}_2 $ cascade decays
through a light chargino are allowed. The physical composition of
$\tilde{\chi}^{\scriptscriptstyle  0}_{1,2} $ is given by a mixture of
comparable components of $\tilde{\gamma} $ and $\tilde{Z} $ with a
small percentage of higgsino components (\hbox{cfr.}{}\
\hbox{Tab.~}{}\ref{tab:scentgb15}). In \hbox{Fig.~}{}\ref{wbrvsm0d},
the only qualitative new feature in the decay pattern is the presence
of sizeable widths for cascade decays. The latter are almost constant
versus $m_0$ and give rise to BR's up to 15\% for the channel
$\tilde{\chi}^{\scriptscriptstyle  0}_2  \rightarrow
\tilde{\chi}^{\scriptscriptstyle  \pm}_1  q \bar{q}^{\prime}$ and up
to 3\% for $\tilde{\chi}^{\scriptscriptstyle  0}_2  \rightarrow
\tilde{\chi}^{\scriptscriptstyle  \pm}_1  e^{\scriptscriptstyle  \pm}
\nu_{\scriptscriptstyle  \ell} $ at $m_0 \simeq 500 {\rm\,GeV} $.

In \hbox{Figs.~}{}\ref{wbrvsm0hm} and \ref{wbrvsm0hp}, we present the
$\tilde{\chi}^{\scriptscriptstyle  0}_2 $ widths and BR's in the High
Cross Section regions. In particular, we consider the scenarios
H$^{\rm \pm}$, defined in \hbox{Tab.~}{}\ref{tab:scentgb15}. One can
check that the higgsino components in these cases are dominant and the
$\tilde{\chi}^{\scriptscriptstyle  0}_1 $ and
$\tilde{\chi}^{\scriptscriptstyle  0}_2 $ are mostly coupled to the
$Z^{\scriptscriptstyle  0} $ boson. Hence, a small dependence on the
scalar masses is found. This is the case especially in the H$^{\rm -}$
scenario (\hbox{cfr.}{}\ \hbox{Fig.~}{}\ref{wbrvsm0hm}), where the BR
for the $q\bar{q} $, $\ell^{\scriptscriptstyle
+}\ell^{\scriptscriptstyle  -} $ and $\nu_{\scriptscriptstyle  \ell}
\bar{\nu}_{\scriptscriptstyle  \ell}  $ channels are quite the same as
for $Z^{\scriptscriptstyle  0} $ decays. In
\hbox{Figs.~}{}\ref{wbrvsm0hm} and \ref{wbrvsm0hp}, the widths and
BR's for the cascade channels are also reported. The relative
importance of these decay modes is considerable only in the H$^{\rm
+}$ case, where the corresponding BR's reach about 18\% for the
hadronic mode, and more than 2\% for each leptonic channel.

We now proceed to the study of the $\tan \beta $ dependence of the
$\tilde{\chi}^{\scriptscriptstyle  0}_2 $ decay pattern. To this end,
as anticipated, we define six new scenarios $\tilde{A}$, $\tilde{B}$,
$\tilde{C}$, $\tilde{D}$, $\tilde{H}^{\pm}$, that are obtained from
the above scenarios by fixing $m_0 = M_{\scriptscriptstyle  Z} $, and
letting free the $\tan \beta $ value. We then study the effect of
changing $\tan \beta $ in the range $1 \div 60$. Note that, although
scenarios A-D were originally defined as lying in the Neutralino
Regions, the variation of $\tan \beta $ can shift such regions above
some of these scenarios (\hbox{cfr.}{}\ \hbox{Fig.~}{}\ref{lep2scen}).
This corresponds to study situations in the $(\mu, M_2)$ plane where
also chargino production is allowed. Furthermore, changing $\tan \beta
$ varies both the mass spectrum and the physical composition of
$\tilde{\chi}^{\scriptscriptstyle  0}_{1,2} $. For instance, in
\hbox{Fig.~}{}\ref{massdiffvstgb}, the
$\tilde{\chi}^{\scriptscriptstyle  0}_2
$-$\tilde{\chi}^{\scriptscriptstyle  0}_1 $ and
$\tilde{\chi}^{\scriptscriptstyle  0}_2
$-$\tilde{\chi}^{\scriptscriptstyle  \pm}_1 $ mass differences (that
are crucial quantities entering the phase-space factor of the
$\tilde{\chi}^{\scriptscriptstyle  0}_2 $ widths) are shown versus
$\tan \beta $, for the scenarios $\tilde{A}$-$\tilde{H}^+$ Some
influence of $\tan \beta $ is also observed in the scalar mass
spectrum (\hbox{cfr.}{}\ Appendix B).

In \hbox{Figs.~}{}\ref{wbrvstgba}--\ref{wbrvstgbhp}, the behaviour of
$\tilde{\chi}^{\scriptscriptstyle  0}_2 $ widths and BR's as functions
of $\tan \beta $ is shown. In the scenario $\tilde{A}$, we can
distinguish three different regimes. For $\tan \beta
\;\raisebox{-.5ex}{\rlap{$\sim$}} \raisebox{.5ex}{$<$}\; 1.3$, the
tree level decay into a light Higgs $h^{\scriptscriptstyle  0} $ is
allowed and the corresponding BR is greater than 80\%. Around $\tan
\beta  = 1.5$, there is a small transition region where the
charged-leptonic channel saturates the BR, since the decay into a
$\tilde{e}_{\scriptscriptstyle  L,R} $ is the only possible two-body
real channel (this corresponds to the scenario A described above). For
larger $\tan \beta $, there is a combined effect of the relative
decreasing of the sneutrino mass with respect to the charged slepton
masses (\hbox{cfr.}{}\ Appendix B) and the increasing Z-ino component
in the $\tilde{\chi}^{\scriptscriptstyle  0}_1 $, that enhances the
$\tilde{\chi}^{\scriptscriptstyle  0}_2  \rightarrow
\tilde{\chi}^{\scriptscriptstyle  0}_1  \nu_{\scriptscriptstyle  \ell}
\bar{\nu}_{\scriptscriptstyle  \ell}  $ decay. Therefore, for large
$\tan \beta $ the $\tilde{\chi}^{\scriptscriptstyle  0}_2 $ gives rise
mostly to missing energy and momentum. In the scenario $\tilde{B}$
(\hbox{Fig.~}{}\ref{wbrvstgbb}), $\tilde{\chi}^{\scriptscriptstyle
0}_2  \rightarrow   \tilde{\chi}^{\scriptscriptstyle  0}_1  q\bar{q} $
is dominant in the whole $\tan \beta $ range considered, although the
charged lepton channel has a considerable BR for $\tan \beta
\;\raisebox{-.5ex}{\rlap{$\sim$}} \raisebox{.5ex}{$>$}\; 5 \div 10$.
In scenario $\tilde{C}$ (\hbox{Fig.~}{}\ref{wbrvstgbc}), the hadronic
channel is the main one for $\tan \beta
\;\raisebox{-.5ex}{\rlap{$\sim$}} \raisebox{.5ex}{$<$}\; 1.5$-2, while
the leptonic channels get comparable to the former at higher $\tan
\beta $. Note that, for $\tan \beta  \;\raisebox{-.5ex}{\rlap{$\sim$}}
\raisebox{.5ex}{$>$}\; 10$, also cascade decays into a
$\tilde{\chi}^{\scriptscriptstyle  \pm}_1 $ give a sizeable
contribution.

As for scenario $\tilde{D}$, we note in \hbox{Fig.~}{}\ref{wbrvstgbd}
a maximum in both the invisible and the hadronic BR's curves
corresponding to a deepening of the charged-leptonic one for $\tan
\beta  \simeq 10$. This is due to the sudden opening of the channel
$\tilde{\chi}^{\scriptscriptstyle  0}_2  \rightarrow
e^{\scriptscriptstyle  \pm}  \tilde{e}^{\scriptscriptstyle
\mp}_{\scriptscriptstyle  L,R} $.

In \hbox{Figs.~}{}\ref{wbrvstgbhm} and \ref{wbrvstgbhp}, we study the
scenarios $\tilde{H}^{\mp}$. Here, we observe again a BR pattern
closely connected to the $Z^{\scriptscriptstyle  0} $ BR's with some
deviation due to the possible presence of a light chargino in the
cascade decays (see also \hbox{Fig.~}{}\ref{massdiffvstgb}). In the
scenario H$^{\rm -}$, the cascade decays contribute considerably at
large $\tan \beta $, while, in the scenario H$^{\rm +}$, they decrease
with $\tan \beta $. Note the strong (although not phenomenologically
relevant) deepening of the radiative decay width at $\tan \beta
\simeq 2.2$, due to destructive interference among various
contributions.

\section{Decreasing the Higgs masses}
\label{sec:Higgs}

In this section we study the sensitivity of the
$\tilde{\chi}^{\scriptscriptstyle  0}_2 $ decay widths and BR's to a
$m_{h^{\scriptscriptstyle  0}} $ change. In particular, we set
$m_{A^{\scriptscriptstyle  0}}  = M_{\scriptscriptstyle  Z} $ which,
compared to the case $m_{A^{\scriptscriptstyle  0}}  = 3
M_{\scriptscriptstyle  Z} $ studied in section 3, corresponds to a
lowering of $m_{h^{\scriptscriptstyle  0}} $ down to 40-70 GeV, at
$\tan \beta  = 1.5$, and to 90-91 GeV, at $\tan \beta  = 30$ in the
considered range of $m_0$ and $M_2$ (\hbox{cfr.}{}\ section 3).

In \hbox{Fig.~}{}\ref{brLH}a, we show the
BR($\tilde{\chi}^{\scriptscriptstyle  0}_2 \rightarrow
\tilde{\chi}^{\scriptscriptstyle  0}_1 h^{\scriptscriptstyle  0} $),
when $m_{A^{\scriptscriptstyle  0}} $ is lowered down to
$M_{\scriptscriptstyle  Z} $. The corresponding reduction of the
threshold for the decay $\tilde{\chi}^{\scriptscriptstyle  0}_2
\rightarrow   \tilde{\chi}^{\scriptscriptstyle  0}_1
h^{\scriptscriptstyle  0} $ considerably extends (with respect to
\hbox{Fig.~}{}\ref{brnlh}) the area where
BR$(\tilde{\chi}^{\scriptscriptstyle  0}_1 h^{\scriptscriptstyle  0} )
> 30\%$ in the $(\mu, M_2)$ plane, down to regions of interest for
LEP2 physics. In particular, this happens in the regions where the
$\tilde{\chi}^{\scriptscriptstyle  0}_{1,2} $ gaugino components are
large, which implies a considerable decrease in the BR's for all the
other decay channels in these regions (as can be checked by comparing
\hbox{Figs.~}{}\ref{brLH}b-d with \hbox{Figs.~}{}\ref{brnee}a,
\ref{brnvv}a and \ref{brnqq}a). On the contrary, the situation is not
altered in the higgsino region, where $m_{\tilde{\chi}^0_1} $ and
$m_{\tilde{\chi}^0_2} $ tend to be degenerate, and the decay
$\tilde{\chi}^{\scriptscriptstyle  0}_2 \rightarrow
\tilde{\chi}^{\scriptscriptstyle  0}_1  h^{\scriptscriptstyle  0} $ is
not allowed.

In \hbox{Figs.~}{}\ref{wbrvsm0a_LH} and \ref{wbrvstgba_LH}, we study
the influence of the $m_{A^{\scriptscriptstyle  0}} $ decrease in the
scenarios $A/\tilde{A}$ considered in section 4. The scenario A is the
only one, out of the six defined in \hbox{Tab.~}{}\ref{lep2scen}, that
falls in the area of large BR($\tilde{\chi}^{\scriptscriptstyle  0}_2
\rightarrow \tilde{\chi}^{\scriptscriptstyle  0}_1
h^{\scriptscriptstyle  0} )$ for $m_{A^{\scriptscriptstyle  0}}  =
M_{\scriptscriptstyle  Z} $. In \hbox{Fig.~}{}\ref{wbrvsm0a_LH}, one
can see the $m_0$ dependence of widths and BR's (already studied in
\hbox{Fig.~}{}\ref{wbrvsm0a} for $m_{A^{\scriptscriptstyle  0}}  =
3M_{\scriptscriptstyle  Z} $), when one sets $m_{A^{\scriptscriptstyle
 0}}  = M_{\scriptscriptstyle  Z} $. The influence of $m_0$ on the
decay width into a Higgs comes from the radiative correction to
$m_{h^{\scriptscriptstyle  0}} $ through the stop mass (\hbox{cfr.}{}\
Appendix B). After the low-$m_0$ range, where the invisible decay
$\tilde{\chi}^{\scriptscriptstyle  0}_2  \rightarrow
\tilde{\nu}_{\scriptscriptstyle  e,L} \nu_{\scriptscriptstyle  e} $ is
still dominant, we have an intermediate range $M_{\scriptscriptstyle
Z}  \;\raisebox{-.5ex}{\rlap{$\sim$}} \raisebox{.5ex}{$<$}\; m_0
\;\raisebox{-.5ex}{\rlap{$\sim$}} \raisebox{.5ex}{$<$}\; 350
{\rm\,GeV} $, where the $b\bar{b}  +\not\!\! E $ signature
(corresponding to the decay into $\tilde{\chi}^{\scriptscriptstyle
0}_1  h^{\scriptscriptstyle  0} $) is largely dominant. After that,
the old pattern of \hbox{Fig.~}{}\ref{wbrvsm0a} is recovered with a
large hadronic BR from $\tilde{\chi}^{\scriptscriptstyle  0}_2
\rightarrow   \tilde{\chi}^{\scriptscriptstyle  0}_1  q\bar{q} $,
through the $Z^{\scriptscriptstyle  0} $-exchange. As for the $\tan
\beta $ dependence (\hbox{Fig.~}{}\ref{wbrvstgba_LH}), the main effect
of lowering $m_{A^{\scriptscriptstyle  0}} $ with respect to
\hbox{Fig.~}{}\ref{wbrvstgba} is an extension of the $\tan \beta $
range where the decay into a light Higgs is relevant, from about
(1-1.4), up to about (1-2). Hence, the $\tan \beta $ range where the
$\tilde{\chi}^{\scriptscriptstyle  0}_2 $ decays in something visible
is quite widened, in the scenario $\tilde{A}$.

\section{The radiative
$\protect\tilde{\chi}^{\scriptscriptstyle  0}_2$ decay}
\label{sec:raddec}

In this section, we study the BR for the decay
$\tilde{\chi}^{\scriptscriptstyle  0}_2  \rightarrow
\tilde{\chi}^{\scriptscriptstyle  0}_1 \gamma$. Provided the mass
difference $(m_{\tilde{\chi}^0_2}  - m_{\tilde{\chi}^0_1} )$ is large
enough as to give rise to a sufficiently energetic photon, this
channel can produce a beautiful signature. A monochromatic photon plus
missing energy and momentum should be observed. Although in all cases
considered in the previous sections,
BR($\tilde{\chi}^{\scriptscriptstyle  0}_2 \rightarrow
\tilde{\chi}^{\scriptscriptstyle  0}_1 \gamma$) never exceeds 15\%, it
can reach values as large as 100\% in particular regions of the $\mu,
M_2, \tan \beta $ space, as we are going to show.

In \hbox{Fig.~}{}\ref{brnph}, the BR($\tilde{\chi}^{\scriptscriptstyle
 0}_2  \rightarrow   \tilde{\chi}^{\scriptscriptstyle  0}_1 \gamma$)
in the $(\mu, M_2)$ plane at fixed $m_0$ and $\tan \beta $ is studied.
We will assume $m_{A^{\scriptscriptstyle  0}}  = 3
M_{\scriptscriptstyle  Z} $ everywhere. Indeed, although
$m_{A^{\scriptscriptstyle  0}} $ sets the mass of the charged Higgs
that flows in the virtual loops (\hbox{cfr.}{}\
\hbox{Fig.~}{}\ref{feyrad}), this parameter is the less critical in
this study. We have checked that varying $m_{A^{\scriptscriptstyle
0}} $ in the range $(M_{\scriptscriptstyle  Z} , 1 {\rm\,TeV} )$ can
change BR($\tilde{\chi}^{\scriptscriptstyle  0}_2 \rightarrow
\tilde{\chi}^{\scriptscriptstyle  0}_1 \gamma$) by at most $\pm 10\%$
(with increasing BR when $m_{A^{\scriptscriptstyle  0}} $ grows). The
scenarios studied in \hbox{Fig.~}{}\ref{brnph} assume either $\tan
\beta  = 1.5$ or 4. The BR for the radiative channel decreases
substantially at larger $\tan \beta $. Furthermore, we set either $m_0
= M_{\scriptscriptstyle  Z} $ or $3 M_{\scriptscriptstyle  Z} $, as in
the previous sections. One can distinguish a specific area where
BR($\tilde{\chi}^{\scriptscriptstyle  0}_2 \rightarrow
\tilde{\chi}^{\scriptscriptstyle  0}_1 \gamma$) is large, that,
particularly for large $m_0$ values (\hbox{cfr.}{}\
\hbox{Fig.~}{}\ref{brnph}c,d), evolves roughly around the $M_2 = - 2
\mu$ line. This corresponds to the region where
$\tilde{\chi}^{\scriptscriptstyle  0}_1 $ is a pure higgsino-B, while
$\tilde{\chi}^{\scriptscriptstyle  0}_2 $ is mostly a photino
(\hbox{cfr.}{}, \hbox{e.g.}{}, \hbox{Figs.~}{} 2.2 and 2.3 in
\hbox{Ref.~}{}\cite{amb-mele}). This situation hinders all the
tree-level decays, since the scalar-exchange decays require gaugino
components in both $\tilde{\chi}^{\scriptscriptstyle  0}_1 $ and
$\tilde{\chi}^{\scriptscriptstyle  0}_2 $, while the
$Z^{\scriptscriptstyle  0} $-exchange ones need higgsino components.
By the way, the different gaugino/higgsino nature of the two lightest
neutralinos also depletes the $\tilde{\chi}^{\scriptscriptstyle  0}_1
\tilde{\chi}^{\scriptscriptstyle  0}_2 $ pair production in
$e^{\scriptscriptstyle  +}e^{\scriptscriptstyle  -} $ collisions. In
\hbox{Fig.~}{}\ref{brnph}, also note that a large fraction of the
high-BR region lies in the area excluded by LEP1 searches.
Nevertheless, one can single out particular situations (of interest
for LEP2 physics and even beyond), where one can have very large BR's
for the radiative channel.

For instance, in \hbox{Fig.~}{}\ref{brvstgbfot}, we study two
different scenarios versus $\tan \beta $: \\ (i)  $\mu = -70
{\rm\,GeV} $, $M_2 = 130 {\rm\,GeV} $, $m_0 = M_{\scriptscriptstyle
Z} , 3M_{\scriptscriptstyle  Z} , 1 {\rm\,TeV} $; \\ (ii) $\mu = -120
{\rm\,GeV} $, $M_2 = 230 {\rm\,GeV} $, $m_0 = M_{\scriptscriptstyle
Z} , 3M_{\scriptscriptstyle  Z} , 1 {\rm\,TeV} $. \\ The scenario (i)
is of interest for LEP2 searches. For instance, for $\tan \beta  =
1.5$, one has $m_{\tilde{\chi}^0_1}  \simeq 65 {\rm\,GeV} $ and
$m_{\tilde{\chi}^0_2}  \simeq 75 {\rm\,GeV} $. The scenario (ii)
concerns heavier neutralino states: for $\tan \beta  = 1.5$, one gets
$m_{\tilde{\chi}^0_1}  \simeq 113 {\rm\,GeV} $ and
$m_{\tilde{\chi}^0_2}  \simeq 127 {\rm\,GeV} $.

Fig. \ref{brnph} shows that BR($\tilde{\chi}^{\scriptscriptstyle  0}_2
\rightarrow  \tilde{\chi}^{\scriptscriptstyle  0}_1 \gamma$) is
particularly large at moderate $\tan \beta $, although the total width
can be as low as a few 10$^{-3}$ eV. The BR is enhanced by increasing
$m_0$, since this makes all the other decays widths decrease further.

On the other hand, one can check that, at $\tan \beta  = 1$,
$\tilde{\chi}^{\scriptscriptstyle  0}_1 $ and
$\tilde{\chi}^{\scriptscriptstyle  0}_2 $ are almost degenerate, while
their mass difference grows monotonically with $\tan \beta $. In order
to have a sufficiently energetic photon, \hbox{e.g.}{}\ $E_{\gamma}
\ge 10 {\rm\,GeV} $, one should restrict to $\tan \beta
\;\raisebox{-.5ex}{\rlap{$\sim$}} \raisebox{.5ex}{$>$}\; 1.5$ in the
case (i) and $\tan \beta  \;\raisebox{-.5ex}{\rlap{$\sim$}}
\raisebox{.5ex}{$>$}\; 1.35$ in the case (ii), which imply
$(m_{\tilde{\chi}^0_2}  - m_{\tilde{\chi}^0_1} )
\;\raisebox{-.5ex}{\rlap{$\sim$}} \raisebox{.5ex}{$>$}\; 10 {\rm\,GeV}
$.

A more in-depth study of the case of large radiative BR will be
carried out in a forthcoming paper \cite{amb-mele2}.

\acknowledgments

We thank Guido Altarelli for constant encouragement and discussions.

\appendix

\section{The neutralino and chargino mass matrices}
\label{appa}

In the MSSM, four fermionic partners of the neutral components of the
SM gauge and Higgs bosons are predicted: the photino $\tilde{\gamma}
$, the Z-ino $\tilde{Z} $ (mixtures of the U(1) $\tilde{B}$ and SU(2)
$\tilde{W}_3$ gauginos), and the two higgsinos
$\tilde{H}^{\scriptscriptstyle  0}_1 $ and
$\tilde{H}^{\scriptscriptstyle  0}_2 $ (partners of the two
Higgs-doublet neutral components). The mixing of these interaction
eigenstates is controlled by a mass matrix $Y$ \cite{neumatr} defined
by:
\begin{mathletters}
\label{eq:neumixing}
\begin{equation}
{\cal L} ^0_M = -\frac{1}{2} \psi^0_i Y_{ij} \psi^0_j + h.c.,
\label{eq:neumasslag}
\end{equation}
where:
\begin{equation}
Y = \left( \begin{array}{cccc}
  M_2 \sin^2\!\theta_{\scriptscriptstyle  W}  +
  M_1 \cos^2\!\theta_{\scriptscriptstyle  W}
& (M_2 - M_1) \sin\theta_{\scriptscriptstyle  W}
  \cos\theta_{\scriptscriptstyle  W}       & 0    & 0    \\
 (M_2 - M_1)\sin\theta_{\scriptscriptstyle  W}
 \cos\theta_{\scriptscriptstyle  W}
& M_2 \cos^2\!\theta_{\scriptscriptstyle  W}
+ M_1 \sin^2\!\theta_{\scriptscriptstyle  W}
& M_{\scriptscriptstyle  Z}   & 0        \\
    0      &      M_{\scriptscriptstyle  Z}
&         \mu \sin 2\beta           &  -\mu \cos 2\beta  \\
    0      &      0       &        -\mu \cos 2\beta
&  -\mu \sin 2\beta     \\
\end{array} \right).
\label{eq:neumat}
\end{equation}
\end{mathletters}
Closely following the notations of \hbox{Refs.~}{}\cite{bartl89,bartl86a},
\hbox{Eqs.~}{}(\ref{eq:neumixing}) are written by suitably choosing the basis:
\begin{equation}
\psi^0_j = (-i\phi_{\gamma}, -i\phi_Z, \psi^a_H, \psi^b_H),
\; \; \; \;  j = 1, \ldots 4,
\label{eq:base}
\end{equation}
where:
\begin{eqnarray*}
\psi^a_H & = & \psi^1_{H_1} \cos \beta  -
\psi^2_{H_2} \sin \beta  \; , \\
\psi^b_H & = & \psi^1_{H_1} \sin \beta  +
\psi^2_{H_2} \cos \beta  \; ,
\end{eqnarray*}
and $\phi_{\gamma}$, $\phi_Z$, $\psi^1_{H_1}$, $\psi^2_{H_2}$ are
two-component spinorial-fields. In \hbox{Eq.~}{}(\ref{eq:neumat}),
$\tan \beta  = \frac{v_{\scriptscriptstyle  2}}{v_{\scriptscriptstyle 1}}$
and $M_{1,2}$ are the U(1)- and
SU(2)-gaugino masses at the EW scale. By assuming gaugino-mass unification
at $M_{\rm \scriptscriptstyle  GUT}$, $M_1$ can be related to $M_2$ by the
equation:
\begin{equation}
M_1 = \frac{5}{3} \tan^2\theta_{\scriptscriptstyle  W} M_2,
\label{eq:unigau}
\end{equation}
that arises from one-loop RGE's (\hbox{cfr.}{}\ Appendix B).
The $Y$ matrix (that, excluding $CP$ violations in this sector of
the model, is real and symmetric) can be diagonalized by a unitary
$4 \times 4$ matrix $N$:
\[ N_{im} N_{kn} Y_{mn} = m_{\tilde{\chi}^0_i}  \delta_{ik}, \]
where $m_{\tilde{\chi}^0_i} $ \ ($i=1, \ldots 4$) is the mass eigenvalue
relative to the $i$-th neutralino state, given by the two-component
spinor field $\chi^{\scriptscriptstyle  0}_i = N_{ij}
\psi^{\scriptscriptstyle  0}_j$.
Then, \hbox{Eq.~}{}(\ref{eq:neumasslag}) can be rewritten, by using
the four-component
neutral Majorana-spinor formalism, in the form:
\[ {\cal L} ^0_M = -\frac{1}{2} \sum_i m_{\tilde{\chi}^0_i}
\bar{\tilde{\chi}^{\scriptscriptstyle  0}_i }
\tilde{\chi}^{\scriptscriptstyle  0}_i , \]
where:
\[ \tilde{\chi}^{\scriptscriptstyle  0}_i  = \left( \begin{array}{c}
                      \chi^0_i          \\
                      \bar{\chi}^0_i    \end{array} \right). \]
The $N$ matrix can be chosen real and orthogonal. In this case some of the
$m_{\tilde{\chi}^0_i} $ eigenvalues can be negative.
The sign of $m_{\tilde{\chi}^0_i} $ is
related to the $CP$ quantum number of the $i$-th neutralino
\cite{hk,bartl86a,petcov}.
By solving a 4-th degree eigenvalue equation, one can find the expressions of
$m_{\tilde{\chi}^0_i} $ and of physical composition of the corresponding
eigenstate in terms of the independent parameter set $\mu$, $M_2$ and
$\tan \beta $ (a complete treatment can be found in
\hbox{Ref.~}{}\cite{bartl89}).

As for the chargino sector,
the corresponding mass term in the Lagrangian is \cite{hk,bartl92}:
\begin{mathletters}
\label{lag:chamass}
\begin{eqnarray}
{\cal L}^{\pm}_M & = & - \frac{1}{2} (\psi^+ \; \psi^-)
\left( \begin{array}{cc} 	0 & X^T \\
	      			X & 0   \\ \end{array} \right)
\left( \begin{array}{c} 	\psi^+ \\
		  		\psi^- \\  \end{array} \right) + h.c. \; ,
					\label{lagchamass} \\
X 		 & = &  \left( \begin{array}{cc}
M_2 	      & M_{\scriptscriptstyle  W}  \sqrt{2}  \sin \beta  \\
M_{\scriptscriptstyle  W}  \sqrt{2}  \cos \beta  & \mu
                               \end{array} \right),    \label{eq:chamat}
\end{eqnarray}
\end{mathletters}
where $\psi^+_j = (-i\phi^+, \; \psi^1_{H_2})$, $\psi^-_j =
(-i \phi^-, \; \psi^2_{H_1})$, \ $j = 1,2$ and $\phi^{\pm}$,
$\psi^1_{H_2}$, $\psi^2_{H_1}$ are two-component spinorial-fields of
W-inos and charged higgsinos, respectively. The mass matrix $X$ can be
diagonalized by two $2 \times 2$ unitary matrices $U$ and $V$:
\[ U_{im} V_{jn} X_{mn} = m_{\tilde{\chi}^{\pm}_i} \delta_{ij} ,  \]
where $m_{\tilde{\chi}^{\pm}_i}$ is the mass eigenvalue for the
$i$-th chargino state, which is defined by:
$\chi^+_i = V_{ij} \psi^+_j$,
$\chi^-_i = U_{ij} \psi^-_j$, \  $i,j = 1,2$ \ ($V$ and $U$ are taken
real after assuming $CP$ conservation).
Here $\chi^{+(-)}_i$ are the two-component spinors corresponding to the
positive- (negative-) charged part of the four-component Dirac-spinor of
$\tilde{\chi}^{\scriptscriptstyle  \pm}_i $. After diagonalization, one
is able to derive a simple formula
for the chargino-mass eigenvalues:
\begin{equation}
m_{\tilde{\chi}^{\pm}_{1,2}}=
\frac{1}{2} \left[\sqrt{(M_2 -\mu)^2+2M_{\scriptscriptstyle  W}^2
(1+\sin 2\beta )} \mp \sqrt{(M_2+\mu)^2
+2M_{\scriptscriptstyle  W} ^2(1-\sin 2\beta )} \right]
\label{charmass}
\end{equation}

We do not consider in this work the small modifications of the above
general scenario that could arise from radiative corrections to
gaugino/higgsino masses. Recent calculations \cite{lahanas93} at the
one-loop level give indication for typical corrections of the order of
6\% (or somewhat higher in particular cases for the lightest
neutralino) with same sign for all neutralino/chargino states. So,
they do not change substantially the relative configuration of
neutralino and chargino masses and do not affect our general
discussion. Also, such corrections are of the same order of magnitude
as other neglected effects, \hbox{e.g.}{}\ other threshold effects in
the RGE evolution.

\section{The scalar-sector mass spectrum}
\label{appb}

In this appendix we collect all relevant formulae we use to calculate
sfermion- and Higgs-mass spectrum in the framework of the MSSM, with
unification assumptions at the GUT scale.

For sfermion masses, once the value of $m_0$ is fixed at the GUT
scale, one finds, by performing the RGE evolution down to the EW scale
\cite{ibanez}:
\begin{equation}
m^2_{\tilde{f}_{L,R}} = \tilde{m}_F^2 + m_f^2 \pm M_D^2,
\label{eq:msf}
\end{equation}
where $m_{\tilde{f}_{L,R}}$ is the mass of the generic sfermion
$\tilde{f}_{L,R}$ and $\tilde{m}_F$, $m_f$ are the corresponding evolved
soft {\sc SuSy}-breaking mass and fermion mass, respectively.
We will name $\tilde{m}_{Q(L)}$ the soft mass for
left squarks (sleptons) and $\tilde{m}_{U_R \ldots E_R}$ the soft masses for
right squarks and charged leptons. In \hbox{Eq.~}{}(\ref{eq:msf}),
$M_D^2$ is the so-called ``D-term'':
\[ M_D^2 = (T_{3,f_{L,R}} - Q_{f_{L,R}}
\sin^2\!\theta_{\scriptscriptstyle  W} ) M_{\scriptscriptstyle  Z} ^2
\cos 2\beta  \; , \]
where $T_{3,f}$ and $Q_f$ are the SU(2)$_{\rm L}$ and U(1)$_{\rm em}$
(in units of $e > 0$) quantum numbers of the fermion $f$.
For the soft masses of the first two generations, Yukawa-coupling effects can
be neglected and simple formulae hold. Indeed, they can be
expressed, as functions of the scale $Q$ and in terms of the common scalar
and gaugino masses $m_0$ and $m_{1/2}$ at the GUT scale
$M_{\rm \scriptscriptstyle  GUT}$
(where $\alpha_1 (M_{\rm \scriptscriptstyle  GUT}) =
\alpha_2 (M_{\rm \scriptscriptstyle  GUT}) =
\alpha_3 (M_{\rm \scriptscriptstyle  GUT}) =
\alpha_{\rm \scriptscriptstyle  GUT} \simeq \frac{1}{25}$),
through the following equations:
\begin{mathletters}
\label{eq:sfmass}
\begin{eqnarray}
\tilde{m}^2_L(t)
& = & m_0^2 + m_{1/2}^2 \frac{\alpha_{\rm \scriptscriptstyle  GUT}}{4 \pi}
\left[ \frac{3}{2} f_2(t) + \frac{3}{10} f_1(t) \right]  , \label{mL} \\
\tilde{m}^2_{E_R}(t)
& = & m_0^2 + m_{1/2}^2 \frac{\alpha_{\rm \scriptscriptstyle  GUT}}{4 \pi}
\left[ \frac{6}{5} f_1(t) \right]                        , \label{mer} \\
\tilde{m}^2_Q(t)
& = & m_0^2 + m_{1/2}^2 \frac{\alpha_{\rm \scriptscriptstyle  GUT}}{4 \pi}
\left[ \frac{8}{3} f_3(t) + \frac{3}{2} f_2(t) + \frac{1}{30} f_1(t) \right]
                                                         , \label{mQ} \\
\tilde{m}^2_{U_R}(t)
& = & m_0^2 + m_{1/2}^2 \frac{\alpha_{\rm \scriptscriptstyle  GUT}}{4 \pi}
\left[ \frac{8}{3} f_3(t) + \frac{8}{15} f_1(t) \right]  , \label{mur} \\
\tilde{m}^2_{D_R}(t)
& = & m_0^2 + m_{1/2}^2 \frac{\alpha_{\rm \scriptscriptstyle  GUT}}{4 \pi}
\left[ \frac{8}{3} f_3(t) + \frac{2}{15} f_1(t) \right]  , \label{mdr}
\end{eqnarray}
\end{mathletters}
where $f_i(t)$ are RGE coefficients at the scale $Q$, given by:
\begin{mathletters}
\label{rge:coeff}
\begin{eqnarray}
f_i(t)  &  =  & \frac{1}{\beta_i} \left( 1 - \frac{1}{(1 + \beta_i t)^2}
\right), \;\;\;\;  i = 1,2,3,                         \label{rge:f} \\
\beta_i &  =  & \frac{b_i}{4\pi}\alpha_{\rm \scriptscriptstyle  GUT},
\;\;\;\;  i = 1,2,3,                                  \label{rge:beta} \\
t       &  =  & \log \frac{M^2_{\rm \scriptscriptstyle  GUT}}{Q^2}.
						     \label{rge:scale}
\end{eqnarray}
\end{mathletters}
In \hbox{Eq.~}{}(\ref{rge:beta}), $b_{1,2,3}$ control the evolution of
U(1), SU(2), SU(3) gauge couplings at the one-loop level.
Assuming for simplicity that the
whole MSSM particle content contributes to the evolution from
$Q \simeq M_{\scriptscriptstyle  Z} $
up to $M_{\rm \scriptscriptstyle  GUT}$, they are:
\begin{equation}
b_i =         \left( \begin{array}{c} b_1 \\ b_2 \\ b_3 \end{array} \right)
    =         \left( \begin{array}{c} 0 \\ -6 \\ -9 \end{array} \right) +
N_{\rm Fam}   \left( \begin{array}{c} 2 \\ 2 \\ 2  \end{array} \right) +
N_{\rm Higgs} \left( \begin{array}{c}3/10 \\ 1/2 \\ 0 \end{array} \right),
\label{bmssm}
\end{equation}
where $N_{\rm Fam} = 3$ is the number of matter supermultiplets and
$N_{\rm Higgs} = 2$ the number of Higgs doublets in the minimal {\sc SuSy}.
Since in the present analysis we use $M_2$ at the EW scale as an independent
parameter in the gaugino sector, we need also the one-loop RGE relation:
\begin{equation}
M_{1,2,3} (M_{\scriptscriptstyle  Z} ) =
\frac{\alpha_{1,2,3} (M_{\scriptscriptstyle  Z} )}
{\alpha_{\rm \scriptscriptstyle  GUT}} m_{1/2}
\Rightarrow
M_3(M_{\scriptscriptstyle  Z} ) =
\frac{\alpha_3(M_{\scriptscriptstyle  Z} )}
{\alpha_2(M_{\scriptscriptstyle  Z} )} M_2(M_{\scriptscriptstyle  Z} )
= \frac{\alpha_3(M_{\scriptscriptstyle  Z} )}
{\alpha_1(M_{\scriptscriptstyle  Z} )} M_1(M_{\scriptscriptstyle  Z} ) \; ,
\label{eq:unigauge}
\end{equation}
which allows us to express $m_{1/2}$ in terms of $M_2$ in
\hbox{Eqs.~}{}(\ref{eq:sfmass}) and from which,
in particular, \hbox{Eq.~}{}(\ref{eq:unigau}) follows.
In order to properly evaluate
the sfermion spectrum through (\ref{eq:sfmass}), we adopt a recursive
procedure (see, \hbox{e.g.}{}, \hbox{Refs.~}{}\cite{kane93,barger}).
First, for any fixed values of $m_0$ and $M_2$, we calculate {\it zero-th
order} sfermion masses $m_{\tilde{f}}^0$ for $Q=M_{\scriptscriptstyle  Z}$,
then we use these values as an input in \hbox{Eqs.~}{}(\ref{eq:sfmass})
(\hbox{i.e.}{}, with $Q=m_{\tilde{f}}^0$ in the corresponding equation for
$\tilde{m}_F$), in order to get out the {\it first order} masses, and so on.
After a few iterations we obtain fast convergence.
In this way, a sufficient agreement with more sophisticated
{\sc SuSy}-spectrum calculations (see, \hbox{e.g.}{},
\hbox{Ref.~}{}\cite{kane93}) is found. In all our analysis, we neglect
both Yukawa-coupling effects in diagonal soft masses and left-right
mixing for the third generation of sfermions.

Concerning the {\sc SuSy}-Higgs sector, starting from the two independent
parameters $m_{A^{\scriptscriptstyle  0}} $ and $\tan \beta $, we calculate
masses from the relations
\cite{ref:higgs}:
\begin{mathletters}
\label{eq:hmass}
\begin{eqnarray}
(m_{H^{\scriptscriptstyle  0}, h^{\scriptscriptstyle  0} })^2
& = & \frac{1}{2} \left[ m_{A^{\scriptscriptstyle  0}} ^2 +
M_{\scriptscriptstyle  Z} ^2 +  \Delta \right]  \nonumber \\
                  & \pm &
\sqrt{ \left[ \left( m_{A^{\scriptscriptstyle  0}} ^2 -
M_{\scriptscriptstyle  Z} ^2 \right) \cos 2\beta  + \Delta \right]^2
 + \left( m_{A^{\scriptscriptstyle  0}} ^2
+ M_{\scriptscriptstyle  Z} ^2 \right)^2 \sin^2 2 \beta},
\label{hneumass} \\
m_{H^{\scriptscriptstyle  \pm}} ^2
& = &  m_{A^{\scriptscriptstyle  0}} ^2 + M_{\scriptscriptstyle  W} ^2,
\label{hchamass}
\end{eqnarray}
\end{mathletters}
where:
\[
\Delta   =
\frac{3}{8\pi^2} \frac{g^2m_t ^4}
{M_{\scriptscriptstyle  W}^2\sin^2\!\beta }
\log \left( 1 + \frac{m_{\tilde{t}} ^2}{m_t ^2} \right) .
\]
As for the mixing angle $\alpha$ in the $CP$-even sector that enters the
$h^{\scriptscriptstyle  0} $ couplings, it is defined at one-loop by:
\begin{equation}
\tan 2 \alpha =
\frac{(m_{A^{\scriptscriptstyle  0}} ^2
+ M_{\scriptscriptstyle  Z}^2) \sin 2\beta }
{(m_{A^{\scriptscriptstyle  0}}^2-M_{\scriptscriptstyle  Z}^2)
\cos 2\beta +\Delta} \; , \:\:\:\:\:\:
- \frac{\pi}{2} < \alpha \le 0 \; .
\label{eq:hmix}
\end{equation}

In our computation, we assume, for the top-quark mass,
$m_t  = 174 {\rm\,GeV} $.
The \hbox{Eqs.~}{}(\ref{hneumass}) and (\ref{eq:hmix})
take into account only the dominant contributions
coming from top/stop loops and we use it under the further assumptions:
$m_{\tilde{t}_{L,R}}  = m_{\tilde{u}_{L,R}}$ and no
$\tilde{t}_{\scriptscriptstyle  L} $-$\tilde{t}_{\scriptscriptstyle  R}$
mixing. All the above simplifications allow us to avoid the introduction
of other {\sc SuSy} parameters, as $A_{\rm \scriptscriptstyle  GUT}$
(or $A_t(M_{\scriptscriptstyle  Z} )$) and $B_{\rm \scriptscriptstyle  GUT}$,
without seriously affecting our results.


\begin{figure}
\caption{Feynman diagrams for the 3-body neutral decays of
the neutralinos.}
\label{fey3bneut}
\end{figure}

\begin{figure}
\caption{Feynman diagrams for the 3-body charged
decays of the neutralinos.}
\label{fey3bcasc}
\end{figure}

\begin{figure}
\caption{Feynman diagrams for the neutralino decays in
neutral Higgses.}
\label{fey2bhiggs}
\end{figure}

\begin{figure}
\caption{Contour plot in the $(\mu, M_2)$ plane for the difference
between the two lightest neutralino masses, for $\tan \beta  = 1.5$ (a)
and 30 (b).}
\label{n2-n1mass}
\end{figure}

\begin{figure}
\caption{Contour plot in the $(\mu, M_2)$ plane for the difference
between the next-to-lightest neutralino mass and the light chargino
mass, for $\tan \beta  = 1.5$ (a) and 30 (b).}
\label{n2-c1mass}
\end{figure}

\begin{figure}
\caption{Feynman diagrams for the radiative neutralino
decay $\protect\tilde{\chi}^{\scriptscriptstyle  0}_2 \rightarrow
\protect\tilde{\chi}^{\scriptscriptstyle  0}_1 \gamma$ in the gauge of
\hbox{Ref.~}{}\protect\cite{hab-wyl}. For each graph shown, there is
a further one with clockwise circulating particles in the loop.}
\label{feyrad}
\end{figure}

\begin{figure}
\caption{Interesting regions and scenarios in the $(\mu, M_2)$ plane
with $\tan \beta  = 1.5$ (a) and 30 (b) for neutralino search at LEP2
($\protect\sqrt{s}  = 190 {\rm\,GeV} $). The NR$^{\rm \pm}$ regions
(bounded by kinematic-limit
curves `N190' and `C190' for
$\protect\tilde{\chi}^{\scriptscriptstyle  0}_1
\protect\tilde{\chi}^{\scriptscriptstyle  0}_2 $ and
$\tilde{\chi}^{\scriptscriptstyle  +}_1
\tilde{\chi}^{\scriptscriptstyle  -}_1 $ production, respectively)
and HCS$^{\rm \pm}$ regions (outlined by the 1 pb contour plot for
the $\protect\tilde{\chi}^{\scriptscriptstyle  0}_1
\protect\tilde{\chi}^{\scriptscriptstyle  0}_2 $
total cross section, for $m_0 = 3M_{\scriptscriptstyle  Z} $) are
indicated. The shaded area corresponds to LEP1 limits.}
\label{lep2scen}
\end{figure}

\begin{figure}
\caption{Contour plot in the $(\mu, M_2)$ plane for the BR (\%) of the
decay $\protect\tilde{\chi}^{\scriptscriptstyle  0}_2
\rightarrow \protect\tilde{\chi}^{\scriptscriptstyle  0}_1
e^{\scriptscriptstyle  +}e^{\scriptscriptstyle  -} $.
The values of $\tan \beta $ and $m_0$ are shown in each case
(\hbox{cfr.}{}\ \hbox{Eqs.~}{}(\protect\ref{br_cases})).
Lines of different style represent contour levels for different values
of the BR (as indicated, these values can change case-by-case).
All results are obtained assuming $m_{A^{\scriptscriptstyle  0}}
= 3 M_{\scriptscriptstyle  Z} $.}
\label{brnee}
\end{figure}

\begin{figure}
\caption{The same as in \hbox{Fig.~}{}\protect\ref{brnee}, but for the decay
$\protect\tilde{\chi}^{\scriptscriptstyle  0}_2  \rightarrow
\sum_\ell\protect\tilde{\chi}^{\scriptscriptstyle  0}_1
\nu_{\scriptscriptstyle  \ell} \bar{\nu}_{\scriptscriptstyle  \ell}  $.}
\label{brnvv}
\end{figure}

\begin{figure}
\caption{The same as in \hbox{Fig.~}{}\protect\ref{brnee}, but for the decay
$\protect\tilde{\chi}^{\scriptscriptstyle  0}_2  \rightarrow
\sum_q\protect\tilde{\chi}^{\scriptscriptstyle  0}_1 q\bar{q} $.}
\label{brnqq}
\end{figure}

\begin{figure}
\caption{The same as in \hbox{Fig.~}{}\protect\ref{brnee}, but for the decay
$\protect\tilde{\chi}^{\scriptscriptstyle  0}_2  \rightarrow
\protect\tilde{\chi}^{\scriptscriptstyle  \pm}_1
\ell^{\scriptscriptstyle  \mp} \nu_{\scriptscriptstyle  \ell} $.}
\label{brclv}
\end{figure}

\begin{figure}
\caption{The same as in \hbox{Fig.~}{}\protect\ref{brnee}, but for the decay
$\protect\tilde{\chi}^{\scriptscriptstyle  0}_2  \rightarrow
\sum_q\protect\tilde{\chi}^{\scriptscriptstyle \pm}_1
q\protect\bar{q}^{\prime}$.}
\label{brcqq}
\end{figure}

\begin{figure}
\caption{Contour plots in the $(\mu, M_2)$ plane for the BR (\%) of the decay
$\protect\tilde{\chi}^{\scriptscriptstyle  0}_2  \rightarrow
\protect\tilde{\chi}^{\scriptscriptstyle  0}_1
h^{\scriptscriptstyle  0} $.
The values of $\tan \beta $ and $m_0$ are shown in each case.
All results are obtained assuming $m_{A^{\scriptscriptstyle  0}}
= 3 M_{\scriptscriptstyle  Z} $.}
\label{brnlh}
\end{figure}

\begin{figure}
\caption{Widths in KeV (left) and BR's in percentage (right)
for all the $\protect\tilde{\chi}^{\scriptscriptstyle  0}_2 $ decays,
as functions of the common scalar mass $m_0$ (GeV), in the scenario A
of \hbox{Tab.~}{}\protect\ref{tab:scentgb15}.
The solid, grey, dashed, dotted lines represent the
charged-leptonic, neutrino (summed over three species), hadronic
(summed over all light flavours), and radiative channels, respectively.
The dot-dashed lines represent the cascade channels. The grey one is
for the leptonic case $\protect\tilde{\chi}^{\scriptscriptstyle  0}_2
\rightarrow  \protect\tilde{\chi}^{\scriptscriptstyle  \pm}_1
e^{\scriptscriptstyle  \mp} \nu_{\scriptscriptstyle  e} $,
the black one is for the hadronic case
$\protect\tilde{\chi}^{\scriptscriptstyle  0}_2 \rightarrow
\sum_q\protect\tilde{\chi}^{\scriptscriptstyle  \pm}_1
q \protect\bar{q}^{\prime}$,
summed over all light flavours.
All results are obtained assuming $m_{A^{\scriptscriptstyle  0}}
= 3 M_{\scriptscriptstyle  Z} $.}
\label{wbrvsm0a}
\end{figure}

\begin{figure}
\caption{The same as in \hbox{Fig.~}{}\protect\ref{wbrvsm0a}, but in
the scenario B.}
\label{wbrvsm0b}
\end{figure}

\begin{figure}
\caption{The same as in \hbox{Fig.~}{}\protect\ref{wbrvsm0a}, but in
the scenario C.}
\label{wbrvsm0c}
\end{figure}

\begin{figure}
\caption{The same as in \hbox{Fig.~}{}\protect\ref{wbrvsm0a}, but in
the scenario D.}
\label{wbrvsm0d}
\end{figure}

\begin{figure}
\caption{The same as in \hbox{Fig.~}{}\protect\ref{wbrvsm0a}, but in
the scenario H$^-$.}
\label{wbrvsm0hm}
\end{figure}

\begin{figure}
\caption{The same as in \hbox{Fig.~}{}\protect\ref{wbrvsm0a}, but in
the scenario H$^+$.}
\label{wbrvsm0hp}
\end{figure}

\begin{figure}
\caption{Difference between the two lightest neutralino masses (left)
and between the next-to-lightest neutralino mass and the light chargino
mass (right), as functions of $\tan \beta $, in the six scenarios
$\protect\tilde{A}$-$\protect\tilde{H}^{+}$.}
\label{massdiffvstgb}
\end{figure}

\begin{figure}
\caption{Widths in KeV (left) and BR's in percentage (right) for all
the $\protect\tilde{\chi}^{\scriptscriptstyle  0}_2 $ decays, as
functions of $\tan \beta $,  in the scenario $\protect\tilde{A}$.
The solid, grey, dashed, dotted lines represent the
charged-leptonic, neutrino (summed over three species), hadronic
(summed over all light flavours), and radiative channels, respectively.
The grey dashed thick line is for the channel
$\protect\tilde{\chi}^{\scriptscriptstyle  0}_2 \rightarrow
\protect\tilde{\chi}^{\scriptscriptstyle  0}_1
h^{\scriptscriptstyle  0} $.
The dot-dashed lines represent the cascade channels. The grey one is
for the leptonic case $\protect\tilde{\chi}^{\scriptscriptstyle  0}_2
\rightarrow  \protect\tilde{\chi}^{\scriptscriptstyle  \pm}_1
e^{\scriptscriptstyle  \mp} \nu_{\scriptscriptstyle  e} $,
the black one is for the hadronic case
$\protect\tilde{\chi}^{\scriptscriptstyle  0}_2 \rightarrow
\sum_q\protect\tilde{\chi}^{\scriptscriptstyle  \pm}_1
q \protect\bar{q}^{\prime}$,
summed over all light flavours.
All results are obtained assuming $m_{A^{\scriptscriptstyle  0}}  =
3 M_{\scriptscriptstyle  Z} $.}
\label{wbrvstgba}
\end{figure}

\begin{figure}
\caption{The same as in \hbox{Fig.~}{}\protect\ref{wbrvstgba}, but in
the scenario $\protect\tilde{B}$.}
\label{wbrvstgbb}
\end{figure}

\begin{figure}
\caption{The same as in \hbox{Fig.~}{}\protect\ref{wbrvstgba}, but in
the scenario $\protect\tilde{C}$.}
\label{wbrvstgbc}
\end{figure}

\begin{figure}
\caption{The same as in \hbox{Fig.~}{}\protect\ref{wbrvstgba}, but in
the scenario $\protect\tilde{D}$.}
\label{wbrvstgbd}
\end{figure}

\begin{figure}
\caption{The same as in \hbox{Fig.~}{}\protect\ref{wbrvstgba}, but in
the scenario $\protect\tilde{H}^{\scriptscriptstyle  -}$.}
\label{wbrvstgbhm}
\end{figure}

\begin{figure}
\caption{The same as in \hbox{Fig.~}{}\protect\ref{wbrvstgba}, but in
the scenario $\protect\tilde{H}^{\scriptscriptstyle  +}$.}
\label{wbrvstgbhp}
\end{figure}

\begin{figure}
\caption{Contour plot on the $(\mu, M_2)$ plane for the BR (\%) of
the decays:
(a): $\protect\tilde{\chi}^{\scriptscriptstyle  0}_2
   \rightarrow  \protect\tilde{\chi}^{\scriptscriptstyle  0}_1
   h^{\scriptscriptstyle  0} $,
(b): $\protect\tilde{\chi}^{\scriptscriptstyle  0}_1
   e^{\scriptscriptstyle  +}e^{\scriptscriptstyle  -} $,
(c): $\sum_\ell\protect\tilde{\chi}^{\scriptscriptstyle  0}_1
   \nu_{\scriptscriptstyle  \ell} \bar{\nu}_{\scriptscriptstyle  \ell}$,
(d): $\sum_q\protect\tilde{\chi}^{\scriptscriptstyle  0}_1 q\bar{q}$,
in the case of a light Higgs ($m_{A^{\scriptscriptstyle  0}}
= M_{\scriptscriptstyle  Z} $), for $m_0 = M_{\scriptscriptstyle  Z}$
and $\tan \beta  = 1.5$.}
\label{brLH}
\end{figure}

\begin{figure}
\caption{Widths in KeV (left) and BR's in percentage (right) for all
$\protect\tilde{\chi}^{\scriptscriptstyle  0}_2 $ decays, as functions
of the common scalar mass $m_0$ in the scenario A.
The solid, grey, dashed, dotted lines represent the
charged-leptonic, neutrino (summed over three species), hadronic
(summed over all light flavours), and radiative channels, respectively.
The dot-dashed line shows only the peaks of the width for the cascade
channel $\protect\tilde{\chi}^{\scriptscriptstyle  0}_2  \rightarrow
\protect\tilde{\chi}^{\scriptscriptstyle  \pm}_1
e^{\scriptscriptstyle  \mp} \nu_{\scriptscriptstyle  e}$
(\hbox{cfr.}{}\ \hbox{Fig.~}{}\protect\ref{wbrvsm0a}).
The grey dashed thick line is for the channel
$\protect\tilde{\chi}^{\scriptscriptstyle  0}_2 \rightarrow
\protect\tilde{\chi}^{\scriptscriptstyle  0}_1
h^{\scriptscriptstyle  0} $.
All results are obtained assuming $m_{A^{\scriptscriptstyle  0}}  =
M_{\scriptscriptstyle  Z} $.}
\label{wbrvsm0a_LH}
\end{figure}

\begin{figure}
\caption{Widths in KeV (left) and BR's in percentage (right) for
all the $\protect\tilde{\chi}^{\scriptscriptstyle  0}_2$ decays,
as functions of $\tan \beta $, in the scenario $\protect\tilde{A}$.
All results are obtained assuming $m_{A^{\scriptscriptstyle  0}}
= M_{\scriptscriptstyle  Z} $.}
\label{wbrvstgba_LH}
\end{figure}

\begin{figure}
\caption{Contour plot in the $(\mu, M_2)$ plane for the BR (\%)
of the radiative decay
$\protect\tilde{\chi}^{\scriptscriptstyle  0}_2  \rightarrow
\protect\tilde{\chi}^{\scriptscriptstyle  0}_1 \gamma$.
The values of $\tan \beta $ and $m_0$ are shown in each case.
All results are obtained assuming $m_{A^{\scriptscriptstyle  0}}
= 3 M_{\scriptscriptstyle  Z} $.}
\label{brnph}
\end{figure}

\begin{figure}
\caption{BR (\%) for the radiative decay
$\protect\tilde{\chi}^{\scriptscriptstyle  0}_2 \rightarrow
\protect\tilde{\chi}^{\scriptscriptstyle  0}_1 \gamma$
as function of $\tan \beta $ in the scenarios (i) (left) and
(ii) (right) given in the text. In each scenario, the behaviour
is given for three different values of the common scalar mass:
$m_0 = M_{\scriptscriptstyle  Z} $ (grey line), $m_0 =
3 M_{\scriptscriptstyle  Z}$ (solid line), $m_0 = 1 {\rm\,TeV}$
(dashed line).
All results are obtained assuming $m_{A^{\scriptscriptstyle  0}}
= 3 M_{\scriptscriptstyle  Z} $.}
\label{brvstgbfot}
\end{figure}

\begin{table}
\caption{Interesting scenarios for neutralino production at LEP2
($\protect\sqrt{s}  \simeq 190$ GeV) in the case $\tan \beta  = 1.5$,
$m_0 = M_{\scriptscriptstyle  Z} $.
Mass eigenvalues for charginos and neutralinos are given, as well as
the sfermion spectrum arising from $m_0 = M_{\scriptscriptstyle  Z}$.
For light neutralinos, the physical composition is reported as well.}
\[ \begin{array}{|c|c||c|c|c|c|c|c|}                       \hline
\multicolumn{8}{|c|}
{\bf Scenarios \;\;\;\;  with \;\;\;\;  \tan \beta  = 1.5} \\ \hline\hline
\multicolumn{2}{|c||}
{\bf Scenario} & {\bf A} & {\bf B} & {\bf C} & {\bf D}
& {\bf H^-} & {\bf H^+}                                    \\ \hline
\multicolumn{2}{|c||}
{(\mu,\: M_2)/M_{\scriptscriptstyle  Z}  \;\;  \rightarrow  }
& (-3, \: 1) & (-1, \: 1) & (-1, \: 1.5) & (3, \: 1.5) & (-0.7, \: 3)
& (1, \: 3)                          			   \\ \hline
\multicolumn{2}{|c||}
{M_1 \; {\rm (GeV)} \;\;\;\;  \rightarrow }
         &  45.7 &  45.7 &  68.6 &   68.6 & 137.2 & 137.2  \\ \hline\hline
         &  {\rm Mass \: (GeV)}
         &  49.5 &  51.5 &  73.7 &   56.0 &  62.3 &  44.9  \\ \cline{2-8}
\tilde{\chi}^{\scriptscriptstyle  0}_1
& (\tilde{\gamma} ,\: \tilde{Z} ) \;\;\;  (\%)
         & (88,\: 11) & (91,\: 6) & (76,\: 10) & (47,\: 45) & ( 0,\:  1)
         & ( 4,\: 20)                                      \\ \cline{2-8}
         & (\tilde{H}^{\scriptscriptstyle  0}_a ,\:
\tilde{H}^{\scriptscriptstyle  0}_b ) \; (\%)
         & ( 1,\:  0) &  (2,\: 2) & ( 1,\: 13) & ( 7,\:  1) & ( 2,\: 97)
         & (70, \: 5)                                      \\ \hline
         & {\rm Mass \: (GeV)}
         & 107.0 &  85.2 &  89.8 &  108.2 & -89.1 & -92.3  \\ \cline{2-8}
\tilde{\chi}^{\scriptscriptstyle  0}_2  & (\tilde{\gamma} ,\:
\tilde{Z} ) \;\;\;  (\%)
         & (12,\: 83) & ( 4,\:  9) & (15,\:  1) & (53,\: 36) & ( 0,\: 7)
         & ( 0,\:  0)                                      \\ \cline{2-8}
         & (\tilde{H}^{\scriptscriptstyle  0}_a ,\:
\tilde{H}^{\scriptscriptstyle  0}_b ) \; (\%)
	 & ( 4,\:  2) & ( 0,\: 86) & ( 2,\: 83) & (10,\:  1) & (90,\: 2)
         & ( 5,\: 94)                                      \\ \hline
\multicolumn{2}{|c||}{\tilde{\chi}^{\scriptscriptstyle  0}_3
\;\;\;\;\;\;\;\;\;\;  \; {\rm Mass \: (GeV)}}
&  275.4 & -129.8 & -124.5 & -274.4 & 144.9 & 153.5        \\ \hline
\multicolumn{2}{|c||}{\tilde{\chi}^{\scriptscriptstyle  0}_4
\;\;\;\;\;\;\;\;\;\;  \; {\rm Mass \: (GeV)}}
& -294.9 &  130.0 &  166.4 &  315.6 & 292.6 & 304.6        \\ \hline
\multicolumn{2}{|c||}{\tilde{\chi}^{\scriptscriptstyle  \pm}_1
\;\;\;\;\;\;\;\;\;\;  \; {\rm Mass \: (GeV)}}
&  106.1 &  104.7 &  110.8 & -101.5 &  80.1 & -62.6        \\ \hline
\multicolumn{2}{|c||}{\tilde{\chi}^{\scriptscriptstyle  \pm}_2
\;\;\;\;\;\;\;\;\;\;  \; {\rm Mass \: (GeV)}}
&  291.2 &  136.2 &  166.2 &  310.0 & 292.2 & 303.5        \\ \hline\hline
\multicolumn{2}{|c||}{\tilde{e}_{\scriptscriptstyle  L} ,
\tilde{e}_{\scriptscriptstyle  R} ,\tilde{\nu}_{\scriptscriptstyle  e,L}
\;\;\;\;\;\;  {\rm Mass \: (GeV)} \;\; } &
\multicolumn{2}{c|}{124, 104, 114} & \multicolumn{2}{c|}{152,115,144} &
\multicolumn{2}{c|}{255,160,250}                            \\ \hline
\multicolumn{2}{|c||}{\tilde{u}_{\scriptscriptstyle  L} ,
\tilde{u}_{\scriptscriptstyle  R}  \;\;\;\;\;\;\;\;  {\rm Mass \: (GeV)}} &
\multicolumn{2}{c|}{285, 277} & \multicolumn{2}{c|}{408,395} &
\multicolumn{2}{c|}{773,746}                                \\ \hline
\multicolumn{2}{|c||}{\tilde{d}_{\scriptscriptstyle  L} ,
\tilde{d}_{\scriptscriptstyle  R}  \;\;\;\;\;\;\;\;  {\rm Mass \: (GeV)}} &
\multicolumn{2}{c|}{289,278} & \multicolumn{2}{c|}{411,395} &
\multicolumn{2}{c|}{774,743}                                \\ \hline
\end{array} \]
\label{tab:scentgb15}
\end{table}


\begin{references}

\bibitem{hk} H.~E.~Haber and G.~L.~Kane,
        Phys. Rep. {\bf 117}, 75 (1985).

\bibitem{amb-mele} S.~Ambrosanio and B.~Mele,
	``Neutralino Production as {\sc SuSy} Discovery
	Process at CERN LEP2'', Preprint ROME1-1094/95,
        to be published in {\it Physical Review} {\bf D}.

\bibitem{bartl} A.~Bartl, H.~Fraas, W.~Majerotto,
        Nucl. Phys. {\bf B278}, 1 (1986);
        Z. Phys. {\bf C41}, 475 (1988).

\bibitem{gun-hab} J.~F.~Gunion, H.~E.~Haber \hbox{et al.}{},
        Int. J. Mod. Phys. {\bf A4}, 1145 (1987);
        J~.F.~Gunion and H.~E.~Haber, Phys. Rev. {\bf D37}, 2515 (1988).

\bibitem{raddec} H.~Komatsu and J.~Kubo,
        Phys. Lett. {\bf 157B}, 90 (1985);
        Nucl. Phys. {\bf B263}, 265 (1986);
        H.~E.~Haber, G.~L.~Kane and M.~Quir\'os,
        Phys. Lett. {\bf 160B}, 297 (1985);
        Nucl. Phys. {\bf B273}, 333 (1986).

\bibitem{hab-wyl} H.~E.~Haber and D.~Wyler,
        Nucl. Phys. {\bf B323}, 267 (1989).

\bibitem{L3} M.~Acciarri \hbox{et al.}{}, L3 Coll.,
	Preprint CERN-PPE/95-14.

\bibitem{bartl89} A.~Bartl, H.~Fraas, W.~Majerotto, N.~Oshimo,
	Phys. Rev. {\bf D40}, 1594 (1989).

\bibitem{amb-mele2} S.~Ambrosanio and B.~Mele, in preparation.

\bibitem{neumatr} J.~Ellis and G.~G.~Ross,
        Phys. Lett. {\bf 117B}, 397 (1982);
	J.~M.~Fr\`ere and G.~L.~Kane,
        Nucl. Phys. {\bf B223}, 331 (1983).

\bibitem{bartl86a} A.~Bartl, H.~Fraas, W.~Majerotto,
        Nucl. Phys. {\bf B278}, 1 (1986);
        A.~Bartl, W.~Majerotto, B.~M\"osslacher, Proc. of
        the Joint Int. Lepton-Photon Symp. \& Europhys. Conf.
        on High Energy Phys., Geneva, Jul. 25-Aug. 1, 1991,
        vol. I, p.357.

\bibitem{petcov} S.~T.~Petcov, Phys. Lett. {\bf 139B}, 421 (1984);
        S.~M.~Bilenky, N.~P.~Nedelcheva and S.~T.~Petcov,
	Nucl. Phys. {\bf B247}, 61 (1984);
        J.~Ellis, J.~M.~Fr\`ere, J.~S.~Hagelin, G.~L.~Kane
        and S.~T.~Petcov, Phys. Lett. {\bf 132B}, 436 (1983).

\bibitem{bartl92} A.~Bartl, H.~Fraas, W.~Majerotto, B.~M\"osslacher,
	Z. Phys. {\bf C55}, 257 (1992).

\bibitem{lahanas93} A.~B.~Lahanas, K.~Tamvakis and N.~D.~Tracas,
        Phys. Lett. {\bf 324B}, 387 (1994);
	D.~Pierce and A.~Papadopoulos,
        Phys. Rev. {\bf D50}, 565 (1994);
        Nucl. Phys. {\bf B430}, 278 (1994).

\bibitem{ibanez} L.~E.~Ib\'a\~nez and C.~L\'opez,
        Nucl. Phys. {\bf B233}, 511 (1984);
	L.~E.~Ib\'a\~nez, C.~L\'opez and C.~Mu\~noz,
        {\it ibid.} {\bf B256}, 218 (1985);
        L.~E.~Ib\'a\~nez and G.~G.~Ross,
        in ``Perspectives on Higgs Physics'', G.~L.~Kane (Ed.),
        p.229 and references therein;
	W.~de Boer, {\it Prog. Part. Nucl. Phys.} {\bf 33}, 201 (1994);
        W.~de Boer, R.~Ehret and D.~I.~Kazakov, Preprint
        IEKP-KA/94-05, hep-ph/9405342 (1994).

\bibitem{kane93} G.~L.~Kane, C.~Kolda, L.~Roszkowski and J.~D.~Wells,
        Phys. Rev. {\bf D49}, 6173 (1994); {\bf D50}, 3498 (1994).

\bibitem{barger} V.~Barger, M.~S.~Berger and P.~Ohmann,
	Phys. Rev. {\bf D47}, 1093 and 2038 (1993);
	V.~Barger, M.~S.~Berger, P.~Ohmann and R.~J.~N.~Phillips,
	Phys. Lett. {\bf 314B}, 351 (1993);
	V.~Barger, M.~S.~Berger and P.~Ohmann,
        Phys. Rev. {\bf D49}, 4908 (1994).

\bibitem{ref:higgs} H.~E.~Haber and R.~Hempfling,
        Phys. Rev. Lett. {\bf 66}, 1815 (1991);
	Y.~Okada, M.~Yamaguchi and T.~Yanagida,
        Prog. Theor. Phys. {\bf 85}, 1 (1991);
	Phys. Lett. {\bf 262B}, 54 (1991);
        J.~Ellis, G.~Ridolfi and F.~Zwirner,
	{\it ibid.} {\bf 257B}, 83 (1991);
        {\bf 262B}, 477 (1991);
        R.~Barbieri, M.~Frigeni and F.~Caravaglios,
        {\it ibid.} {\bf 258B}, 167 (1991);
        A.~Yamada, {\it ibid.} {\bf 263B}, 233 (1991).

\end{references}
\end{document}